\documentclass[useAMS,usenatbib,preprint2]{aastex}
\usepackage{amsmath}
\usepackage{tabularx}
\usepackage{array}

\usepackage{appendix}

\usepackage{colortbl}
\usepackage{rotating}  

\begin{document}

\title{Vast  planes of  satellites in  a high  resolution simulation  of the
  Local Group: comparison to Andromeda}

\author{N. Gillet\altaffilmark{1*}, P. Ocvirk\altaffilmark{1}, D. Aubert\altaffilmark{1}, 
A. Knebe\altaffilmark{2}, N. Libeskind\altaffilmark{3}, 
G. Yepes\altaffilmark{2}, S. Gottl\"ober\altaffilmark{3} \\ 
and Y. Hoffman\altaffilmark{4}}

\affil{$^1$Observatoire astronomique de Strasbourg, Universit\'e de Strasbourg, CNRS, UMR 7550, 11 rue de lUniversit\'e, F-67000 Strasbourg, France}
\affil{$^2$Departamento de F\'{\i}sica Te\'orica, M\'odulo , Universidad Aut\'onomade Madrid, Cantoblanco, E-28049, Spain}
\affil{$^3$Leibniz-Institute f\"ur Astrophysik Potsdam (AIP), An der Sternwarte 16, D-14482 Potsdam, Germany}
\affil{$^4$Racah Institute of Physics, Hebrew University, Jerusalem 91904, Israel}
\email{$^*$E-mail: nicolas.gillet@astro.u
nistra.fr}

\begin{abstract} 

We search for vast planes of satellites (VPoS) in a high resolution
simulation of the Local Group performed by the CLUES project, which improves
significantly the resolution of former similar studies. We use a simple
method for detecting planar configurations of satellites, and validate it on
the known plane of M31.
We implement a range of prescriptions for modelling the satellite
populations, roughly reproducing the variety of recipes used in the
literature, and investigate the occurence and properties of planar
structures in these populations.
The structure of the simulated satellite systems is strongly non-random and
contains planes of satellites, predominantly co-rotating, with, in some
cases, sizes comparable to the plane observed in M31 by Ibata et
al.. However the latter is slightly richer in satellites, slightly thinner
and has stronger co-rotation, which makes it stand out as overall more
exceptional than the simulated planes, when compared to a random population.
Although the simulated planes we find are generally dominated by one real structure, 
forming its backbone, they are also partly fortuitous and are thus 
not kinematically coherent structures as a whole.
Provided that the simulated and observed planes of satellites
are indeed of the same nature, our results suggest that the VPoS of M31 is
not a coherent disc and that one third to one half of its satellites must
have large proper motions perpendicular to the plane.

\end{abstract}

\section{Introduction}
\label{Introduction}

The discovery of the planar distributions of satellite galaxies around the
Milky Way \citep{lynden1976, kunkel1976} and Andromeda \citep{koch2006, McCo2006,
  Ibata2013, conn2013} is regarded as a new
challenge to galaxy formation theory in the context of the standard model of
cosmology $\Lambda$CDM \citep{kroupa2005}.
Using PAndAS (Pan-Andromeda Archaeological Survey), 
\citet{Ibata2013} (hereafter I13) and \citet{conn2013} found that among the 27 
known satellites of Andromeda, 15 are located within a very thin, extended plane 
(with a thickness of 12.6 $\pm$ 0.6 kpc and about 200 kpc in radius). Moreover, 
they estimate, from the radial velocities, that 13 are co-rotating. 
\cite{shaya2013} find that of the 12 remaining satellites, 8 sit on a 
second plane roughly parallel to that found by I13.
While such planar distributions of satellites are not impossible to find in
$\mathrm{\Lambda CDM}$ simulations \citep{aub2004,lib2005, kang2005,zen2005,
  lib2007, deason2011, lib2009}, their frequency and quantitative
resemblance with the observed I13 VPoS is hotly debated. 
\citet{bahl2013} investigated the incidence of planar alignments of
satellite galaxies in the Millennium-II simulation and concluded that vast,
thin planes of dwarf galaxies, similar to that observed in the Andromeda
galaxy (M31), occur frequently in $\mathrm{\Lambda CDM}$ cosmology. Shortly
afterwards, \citet{ibata2014, pawlow2014} re-examined this simulation,
accounting for the observed plane's extent, thickness and abundance, and
came to the opposite conclusion, that only 0.04$\%$ of galaxies possess
planes as extreme as M31's.
These studies were performed, {\em ``with the caveat that the Millennium-II
  simulation may not have sufficient mass resolution to identify confidently
  simulacra of low-luminosity dwarf galaxies''}, as duly noted by
\cite{ibata2014}: the semi-analytic modeling of Guo et al. (2013) differentiates
normal galaxies from ``orphans'', the latter being systems whose parent
sub-halo is no longer resolved. It is possible that many of these orphans
are tidally disrupted, and hence that they are not directly comparable to
the observed dwarf galaxies.
In the present paper, we avoid this caveat by using a high resolution of the
local group performed by the CLUES project, offering an improvement of a
factor 15 in mass resolution with respect to the Millenium-II simulation,
which allows us to resolve the satellites in the mass range of interest more
consistently. This improvement comes however at the cost of volume, as we
are left with only 2 host galaxies to study in the present paper.
In Sec. 2 we present the simulation, the satellite population models used
and the method for detecting planes of satellites. In Sec. \ref{Planes
  detection} we present the results and the detected planes, followed by a
short discussion and our conclusions.

\section{Methodology}
This section describes the simulation used, the satellite population models
and the method for detecting and quantifying satellite alignments.
\label{Selection of satellites}

\subsection{The CLUES simulation}
\label{The CLUES simulation}

The simulation we use in this study was performed by the CLUES (Constrained
Local UniversE Simulation) project 
\citep{clues, yepes2014}, using  GADGET2 \citep{springel2005}.
It was run using standard $\mathrm{\Lambda CDM}$ initial conditions assuming
a WMAP3 cosmology, 
i.e. $\Omega_{m}=0.24$, $\Omega_{b}=0.042$, $\Omega_{\Lambda}=0.76$
\citep{wmap}, and uses a zoom technique, where a small, high resolution
region is embedded in a larger, lowresolution box providing the large scale
cosmological context. The zoom region is about $\mathrm{2h^{-1}Mpc}$ wide at
{\em z}=0 and contains a Local Group analog, with a mass resolution of
$\mathrm{m_{dm}=2.1\times10^{5}h^{-1}M_{\odot}}$ for the high resolution
dark matter particles and $\mathrm{m_{gas}=4.42\times10^{4}h^{-1}M_{\odot}}$
for the gas. The feedback and star formation prescriptions of
\cite{springel2003} were used. For more detail we refer the reader to
\citep{clues}. 
This simulation has been used to investigate a number of properties of
galaxy formation at high resolution 
\citep{jaime2011,knebe2011b,knebe2011a,libeskind2011a,libeskind2011b} and
reionization studies \citep{ocvirk2013,ocvirk2014}. Besides being a
well-studied simulation, the advantage of this dataset for
the present study is twofold.
First of all, it produces a fairly realistic Local Group at {\em z}=0: the
  MW and M31 are in the correct range of separation 
and total virial mass: 5.71 $\times 10^{11} M_{\odot}$ for the MW and $7.81
\times 10^{11} M_{\odot}$ for M31 at a virial 
radius of 220.4 kpc and 244.58 kpc, respectively (Tab.2 of \cite{lib2010}).
Also a cluster of roughly the size of Virgo is found some 12 Mpc 
away from the simulated LG.
In the rest of the paper, and for the sake of clarity, we will refer to the
simulated galaxies as respectively LGa and 
LGb, while MW and the M31 will refer to the real galaxies. 
Secondly, its mass resolution in the zoomed region allows us to resolve 
$\mathrm{M_{halo}=4.2\times10^{6}h^{-1}M_{\odot}}$ haloes, which is
comfortably below the expected or measured mass 
range of the satellite population of M31, and 15 times smaller than the
20-particles haloes of the 
Millenium-II simulation. 
The dark matter catalog are produced by Amiga halo finder \citep{AHF2004, 
AHF2009}, and no haloes that are contaminated with low resolution particles 
are found within the volume considered here and thus only haloes fully resolved by 
the lowest mass particles are used in our analysis.
These {\em z}=0 halo catalogs give the mass, positions and velocities of the dark
matter haloes. 
They will be used to analyse the properties of the satellite populations
obtained.
The simulation was also post-processed with the radiative transfer code ATON
\citep{aubert2008,aubert2010} in order 
to compute a reionization redshift for each halo, which will be used in our
satellite population models. 
This is described in detail in \citep{ocvirk2014}, which also used the
results of this post-processing to study the 
correlation between present-day satellite positions and their reionization
histories.

\subsection{Spatial selection}
\label{Volume of selection}

In order to be able to make the comparison of the simulation with the
observed plane of satellites as direct as possible, 
we first perform a spatial selection of our halo populations as close as
possible to the PAndAS volume. 
We also explore a different, slightly wider volume, and finally consider a
spherical volume:

\begin{itemize}
\item{PAndAS: our first volume is a PAndAS-like volume around the host 
galaxy.
The line of sight is taken along the line linking LGa to LGb, and we fixed
the distance of the observer at 780kpc.
The galaxy LGa will be the observer of LGb and vice-versa. 
The PAndAS area is modeled by a circular area of 12 degrees around Andromeda
and because of the contamination due to Andromeda's stellar disc, satellites
in the central 2.5 degrees are rejected. 
Also detected satellites are constrained to 500kpc from the host forward and
backward. 
We do not consider the extension of the survey around M33, which however
contains two satellites in the observations.}
\item{PAndAS-bis: here we will consider a modified PAndAS volume.
The distance of the observer is increased to 1200 kpc, the outer angle
limited to 10 degrees and the inner to 2 degrees. 
This volume, larger than the original PAndAS volume, allows us to probe
other configurations of the satellite population.}
\item{Spherical: our third volume is a simple sphere around the host, and
  therefore there is no line of sight. Satellites have to be closer than 500
  kpc and further than 50kpc from the host. This type of volume is
  similar to that chosen by \cite{bahl2013} using the Millenium-II
  simulation, which is why we include it in this study.
This simple volume also allows us to explore the spatial distribution of
satellites of the simulation more systematically, without the possible bias
induced by the line of sight.}
\end{itemize}

\par We now turn to modelling the satellite population of our simulated
galaxies.

\subsection{Satellite population models}
\label{Criteria of selection}

The PAndAS survey only detected the 27 brightest satellites, while our
simulation counts thousands of dark matter haloes around each
galaxy. Therefore we need to find out which of our dark matter haloes will
be the 27 brightest, i.e. which ones will have the largest stellar
mass. There is no real consensus on what shapes the properties of satellite
populations, and as a result their modeling is still very
uncertain. Therefore we chose to explore a number of simple recipes, in an
attempt to emulate at least partially the variety of models found in the
literature. 
The initial halo catalog we used gives us for a sphere of 2Mpc around LGa
and LGb, 5563 satellites and 6823 respectively, with their positions,
velocities, dark matter masses at {\em z}=0, maximum dark matter masses throughout
their assembly history $\mathrm{M_{max}}$, stellar masses
$\mathrm{M_{star}}$ and reionization redshift ${\rm z_{r}}$.
We use this data and simple selections on these quantities or combinations
thereof, in order to mimic the basic behaviour of a number of popular
models.

First of all, we consider as surviving haloes at {\em z}=0 only those having
retained more than $5\%$ of their maximal mass $\mathrm{M_{max}}$. 
Haloes with larger mass loss are assumed to have experienced strong tidal
disrution during their accretion on the host 
LGa or LGb and have lost their stars to the stellar halo of the host. 
This is similar to what is found in the literature
\citep{busha2010,ocvirk2011,maccio2010}. 
Tests indicate that this criterion does not have a strong impact on our
results. 
We also keep as satellites only those sub-haloes which are gravitationally
bound to the host. 
Then we select the  $N_{sat}$ brightest haloes (we will consider samples of
$N_{sat}$= 25, 27, 30, 35, 50, 100 and 150 
satellites) according to simple stellar content modelling using physically
motivated criterions. 
The five models we considered are listed below. 
We do not focus on the absolute stellar mass content given by these models,
but only use them in a relative manner, so as to determine the $N_{sat}$
brightests.

\begin{itemize}
\item{{\bf $\mathrm{M_{star}}$}: the CLUES hydrodynamical simulation we used
  spawns star particles using the \cite{springel2003} formalism. Therefore a
  stellar mass $\mathrm{M_{star}}$ can be computed for all dark matter
  haloes. However, the properties of low mass satellites populations are
  notoriously difficult to reproduce even with high resolution
  hydro-dynamical runs. Moreover, the simulation used a uniform UV
  background at reionization and therefore does not account for local
  inside-out effects such as shown in \citet{ocvirk2013,
    ocvirk2014}. Therefore we decided to explore several modelling
  alternatives.}
\item{ {\bf ${\rm M_{z=0}}$} (simple abundance matching): here we assume 
that the brightest satellites should be the most massive ones at {\em z}=0. 
This is the basic underlying assumption of the abundance matching technique, 
widely used in semi-analytical modelling. This assumption is supported 
by the results of e.g. \cite{brook2014}, where the stellar mass is taken to be 
a monotonous function of halo mass at z=0, but challenged by other groups 
\citep{sawala2014} because of the stochasticity of star formation at the mass 
scale of the faint M31 and Milky Way satellites.}
\item{{\bf $\mathrm{z_{r}}$} (reionization reshifts): For each satellite we
  computed the redshift of last reionization using the results of the
  radiative transfer post-processing of \cite{ocvirk2014}. Reionization is
  thought to be one of the main causes of the low-efficiency of star
  formation in low mass satellites, as suggested for instance by
  \cite{brown2014}. Very often semi-analytical models account for
  reionization by shutting down star formation in low mass haloes
  \citep{koposov2008,busha2010,ocvirk2011} at $\mathrm{z_{r}}$. Therefore
  one could expect that the haloes with the latest reionization redshift
  $\mathrm{z_{r}}$ will be the brightest.}
\item{ {\bf $\mathrm{M_{max}}$}: according to \cite{gnedin2000}, the effect
  of reionization on the baryonic fraction inside dark matter haloes is a
  function of mass, and the transition between sterile and UV-immune haloes
  takes place over 2 decades in mass. Therefore one could expect that the
  total stellar mass of a satellite progenitor halo is mostly tied to the
  maximum mass $\mathrm{M_{max}}$ it has been able to reach throughout its
  life. Under this assumption the brightest satellites would be the one with
  the larger $\mathrm{M_{max}}$, even if they are not the most massive at
  {\em z}=0. This is similar to the assumptions of \cite{most2013} for satellite galaxies.}
\item{{\bf $\mathrm{z_{r}M_{max}}$}: the last model is an attempt at
  accounting for the mass scale at which haloes become UV-immune, {\em} and
  the variety of reionization histories of lower mass haloes. For instance,
  \citep{paw2013} showed that haloes more massive than $10^{9}M_{\odot}$ are
  insensitive to ionizing radiation. Therefore all haloes which have grown
  beyond this threshold must have stars. On the other hand, the remaining
  less massive haloes have seen their star formation history truncated at
  reionization. Under these assumptions, we select the brightest satellites
  as the haloes with $\mathrm{z_{r}M_{max}}>10^{9}M_{\odot}$ (these are 10
  for LGa and 9 for LGb), completed with the satellites with latest
  $\mathrm{z_{r}}$ to obtain a sample of  $N_{sat}$ satellites. 
}

\end{itemize}

\begin{figure*}[ht]
\begin{center}
\includegraphics[scale=0.305]{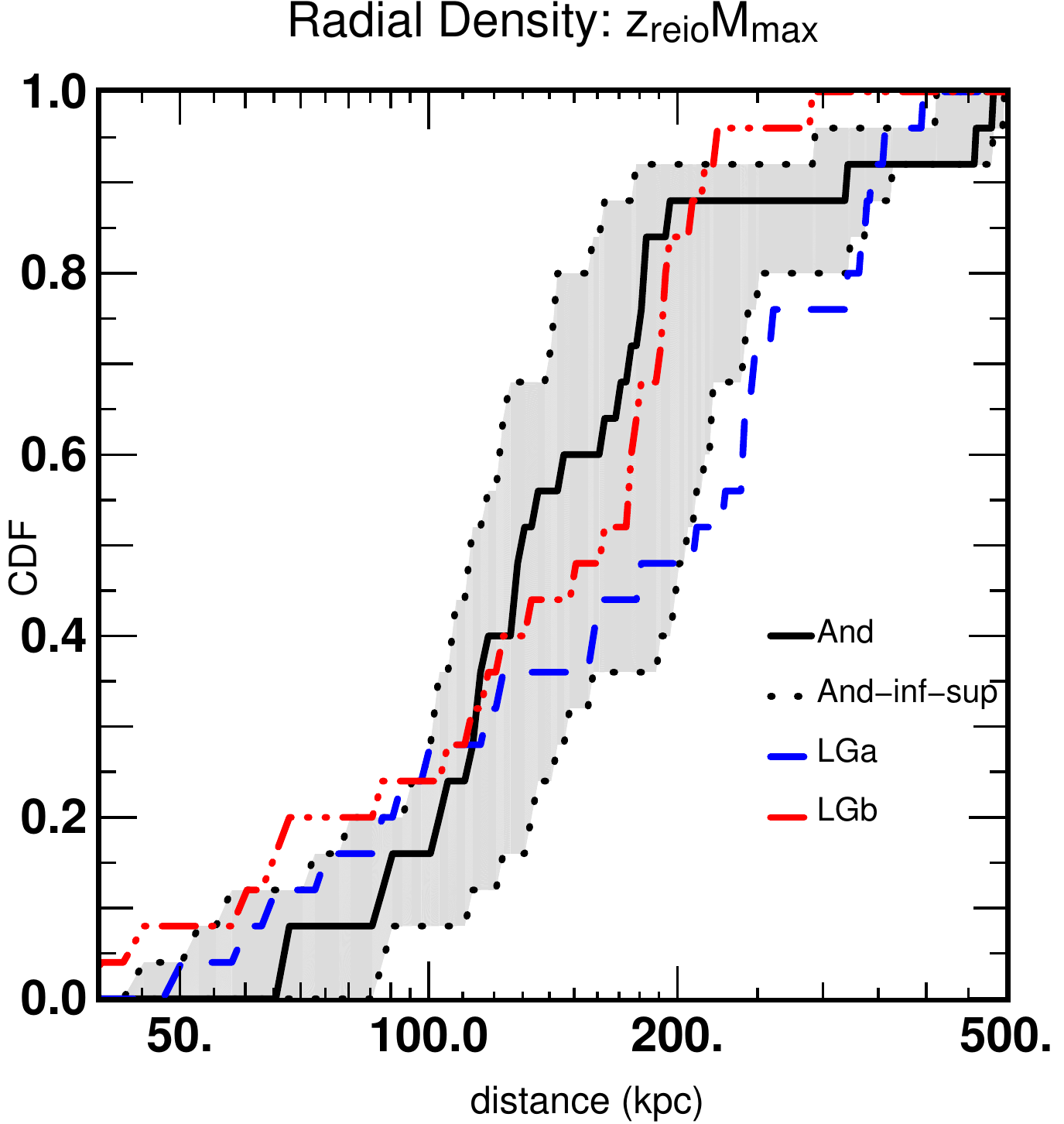}
\includegraphics[scale=0.305]{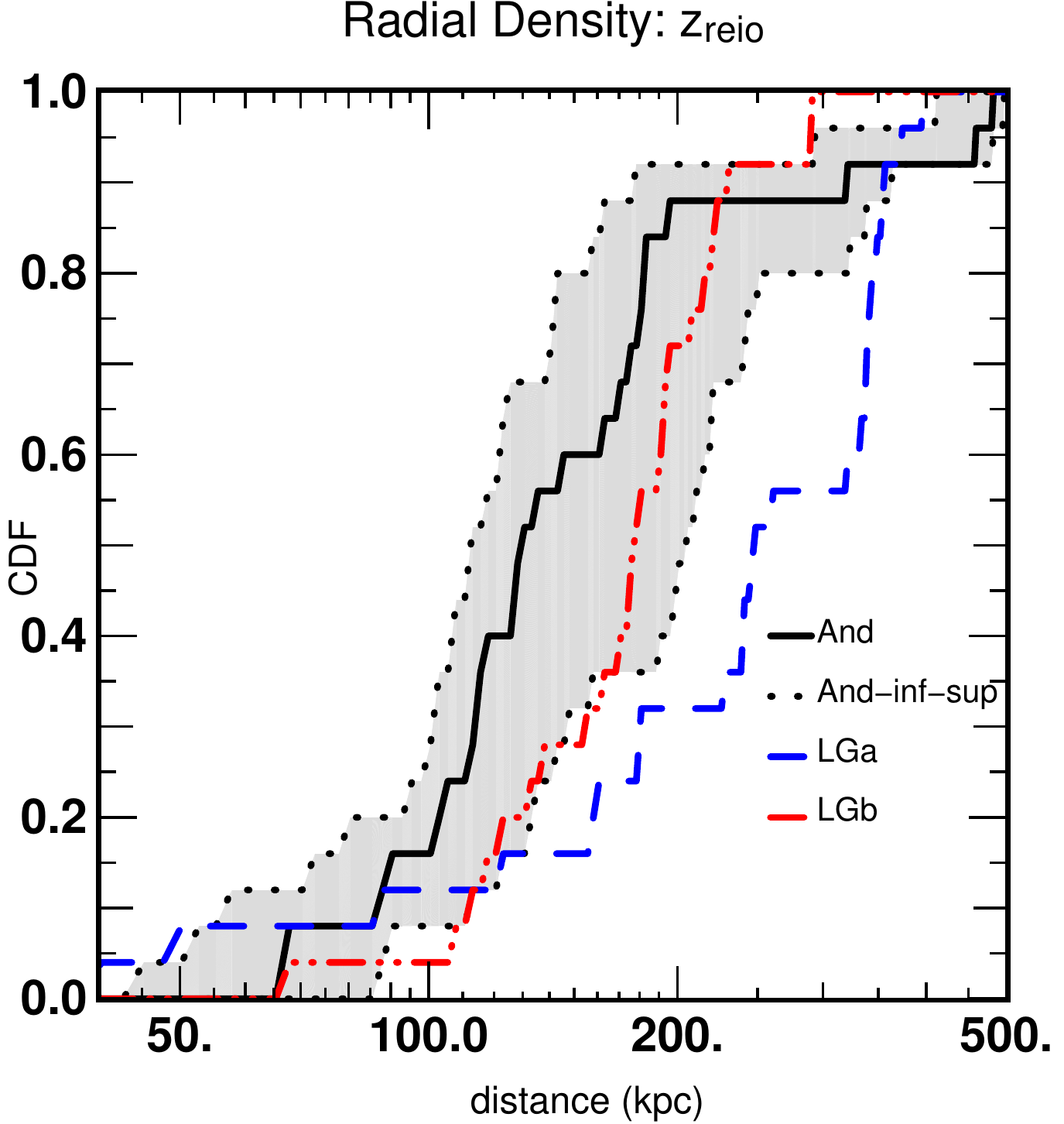}
\includegraphics[scale=0.305]{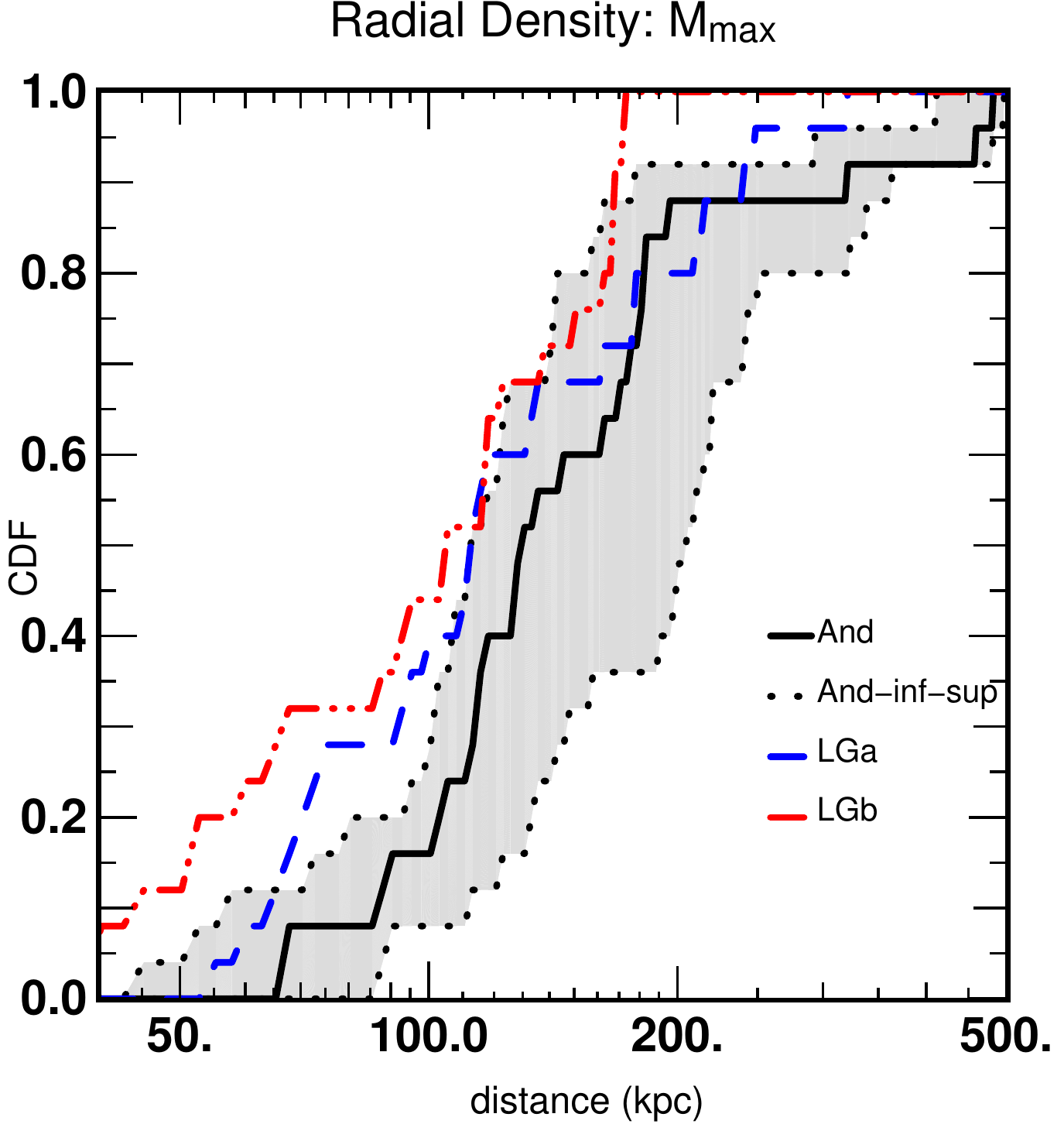}
\includegraphics[scale=0.305]{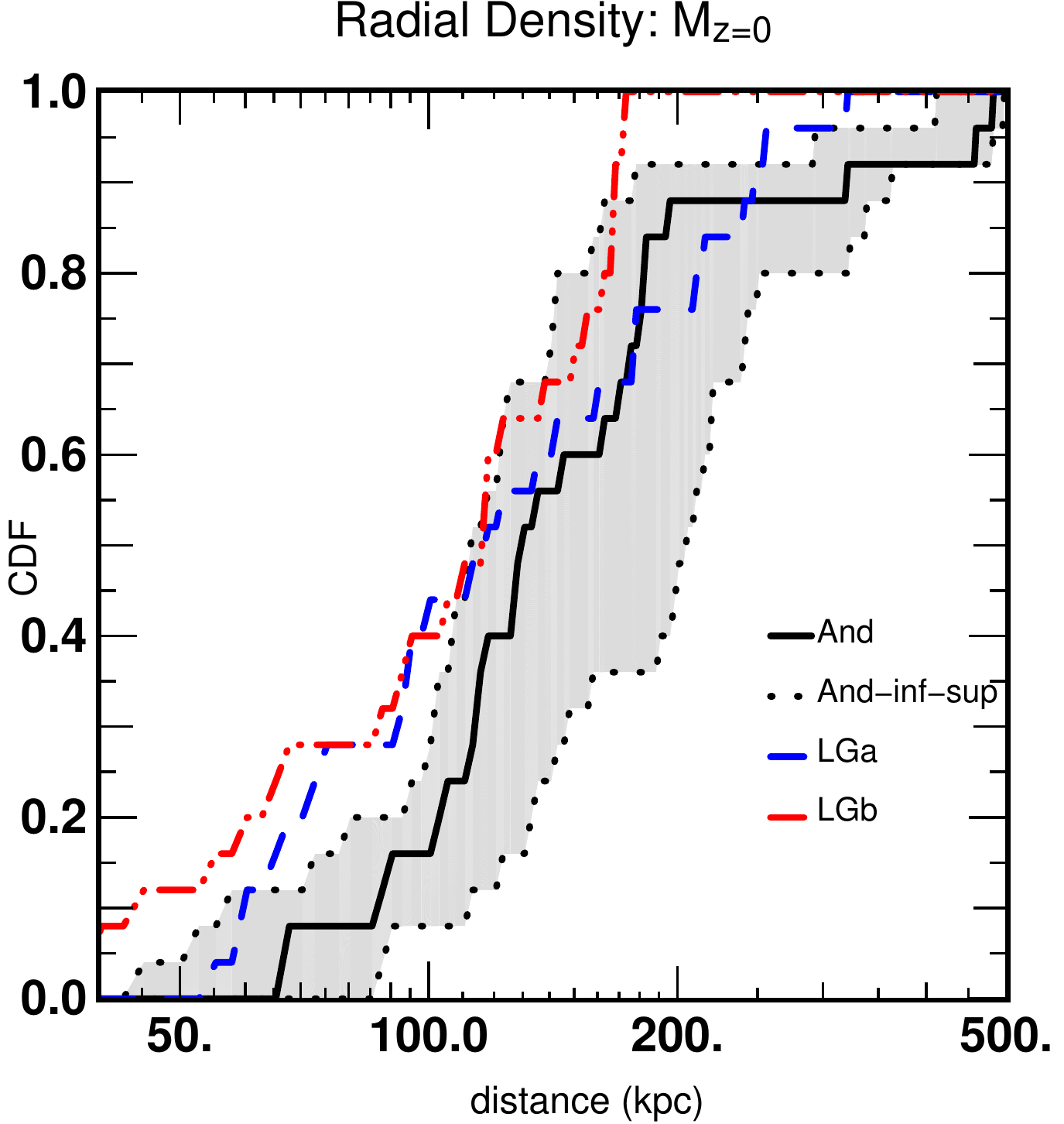}
\includegraphics[scale=0.305]{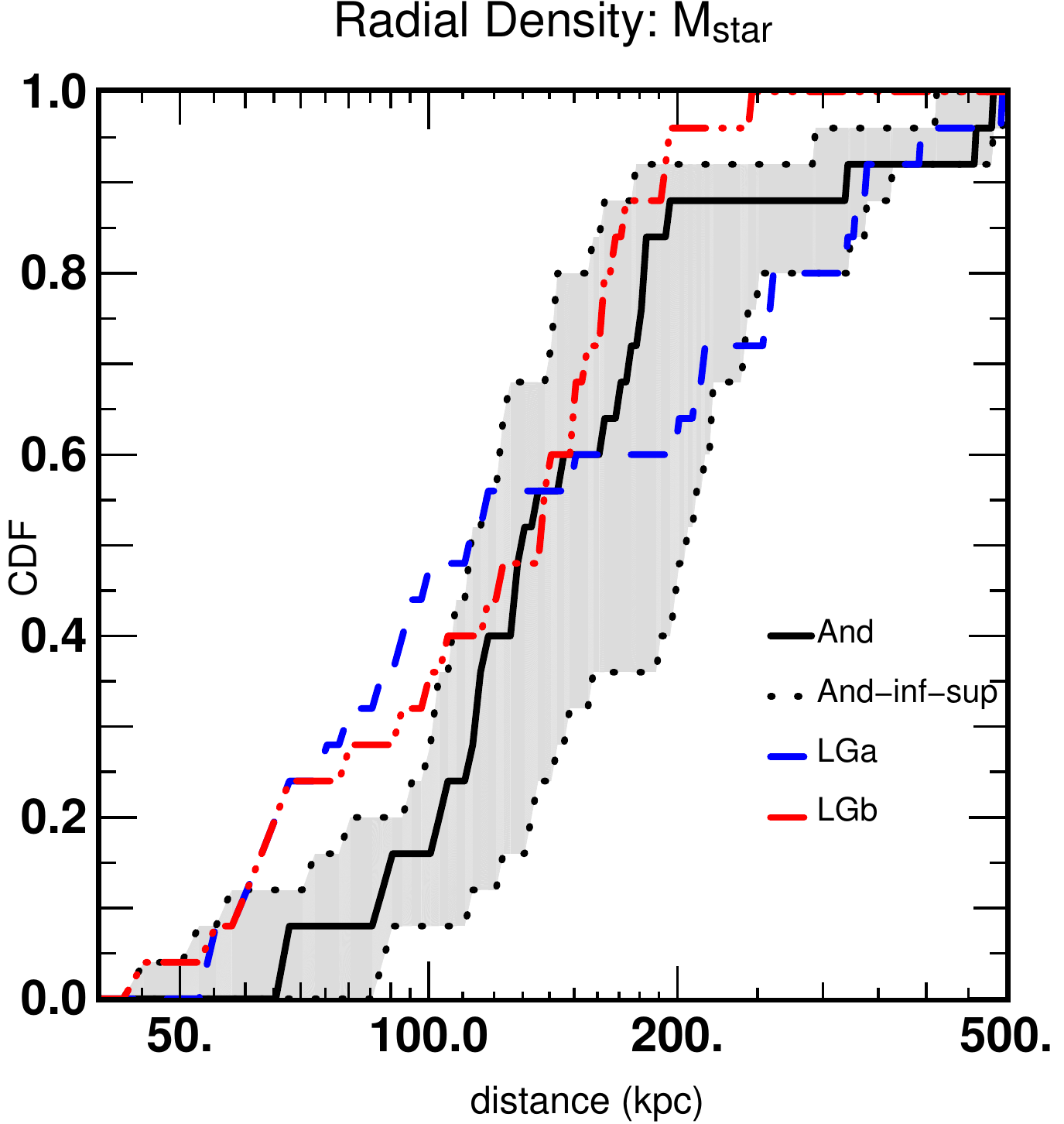}
\includegraphics[scale=0.305]{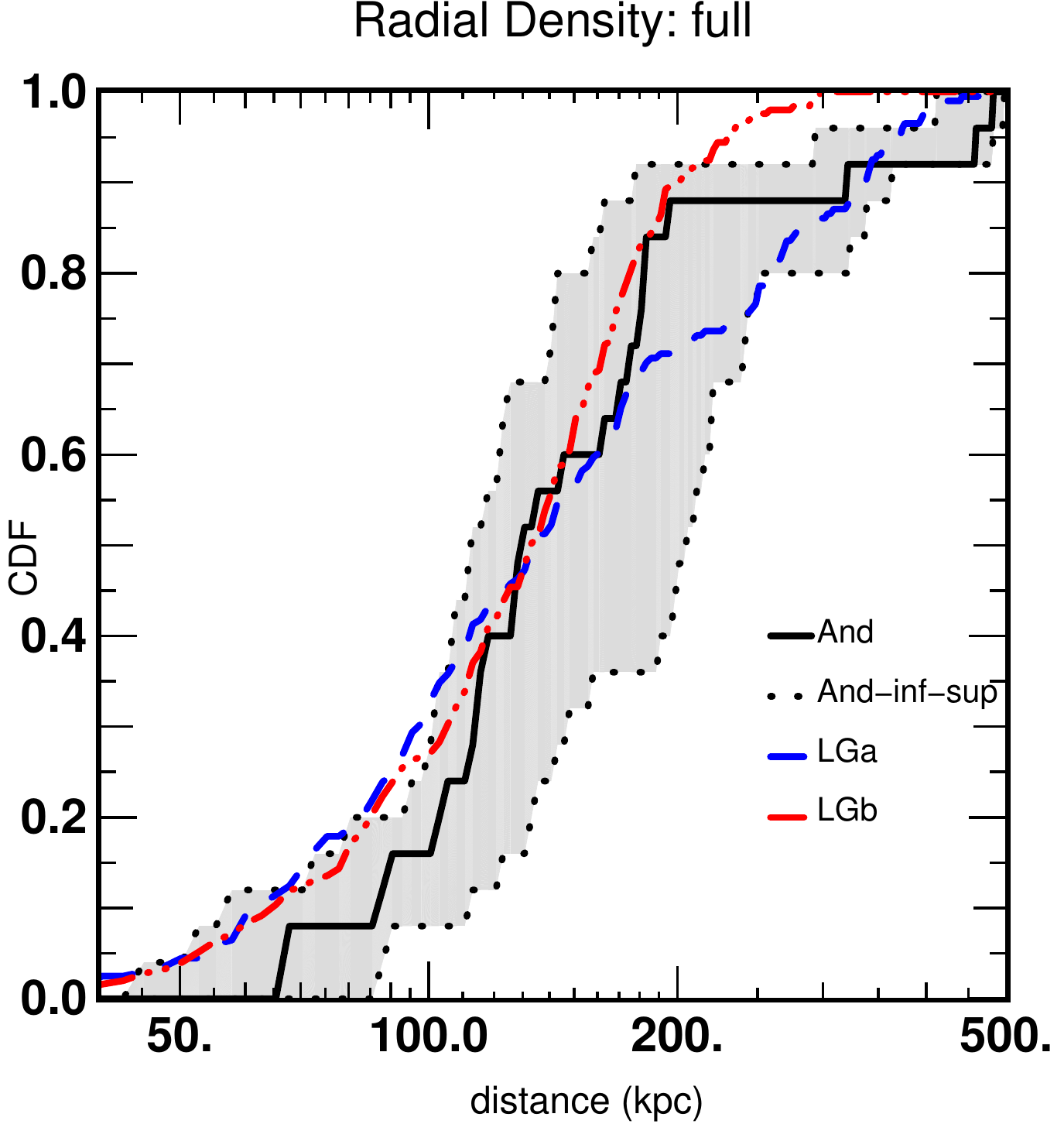}
\caption{\small\label{fig_SDP}
Comparison of the radial distribution observed around Andromeda (black) with
the simulated galaxies LGa and LGb (respectively 
dashed blue and double dot dashed red). 
The comparison is done for the five models, within a PAndAS volume
including 25 satellites. 
The one-$\sigma$ uncertainties of the observations are the gray area. 
They are computed from the two extreme cases, where all the satellites are
at the lower distance or the upper one (dot black curves). 
The {\em bottom right} panel represent the radial distribution for the
full sample of satellite haloes, without any selection.
It contains 245 satellites for LGa and 326 for LGb. 
}
\end{center}
\end{figure*}

We do not focus on the absolute stellar mass content given by these models,
but only use them in a relative manner, 
so as to determine the $N_{sat}$ brightests. 
This modelling does not yield independent populations. Indeed, they will
have some fraction of their satellites in 
common, depending on the number of bright satellites $N_{sat}$ considered. 
Here we do not try to tune our models for reproducing the properties of the
observed population other than their number. 
This is notoriously complex, and beyond the scope of this paper. 
Instead, their rather different outcomes demonstrate the range of properties
allowed for the model population. 
For instance, Fig. \ref{fig_SDP} shows the radial distributions obtained
within a PAndAS volume for our 5 models and 
the whole population of dark matter haloes, for LGa and LGb. 
First of all, we note that the two galaxies exhibit different satellite
populations. 
Indeed the LGa satellite system is, in most of the cases, more extended than
LGb's. 
Besides, all models relying on ${\rm M_{star}}$, ${\rm M_{z=0}}$ or
${\rm M_{max}}$ are too concentrated, while 
the ${\rm z_{r}}$ model is too extended, because of the typical inside-out
reionization process described 
in \cite{ocvirk2011}, \cite{ocvirk2013} and \cite{ocvirk2014}: a late
$\rm{z_{r}}$ selection yields more remote haloes. 
Finally, the ${\rm z_{r}M_{max}}$ model, while doing slightly better, is
still is not a great fit to the observations. 
Even without a rigorous Kolmogorov-Smirnov test, it is clear that none of
our models reproduces the observed distribution very well. 
This misfit is not necessarily caused by our modelling: fairly large
 variations in the radial distribution of a galaxy of a given mass are
 expected due to cosmic variance, as a result of different mass assembly
 histories.For instance, \cite{lun2012} shows, using the Aquarius simulations, 
 that the radius containing half of the satellites can vary from 50kpc to 120kpc 
 within the 6 MW realizations of the dataset.
We recall however that constrained simulations such as the one we used here exhibit 
smaller variance than the baseline cosmic variance \cite{jaime2011}. How this affects the 
radial distribution of satellites is unclear.
The radial distribution of the MW's and M31's satellites are also unlike: 
 although they are similar within 100 kpc, they differ dramatically in the outskirts 
 of the halo \citep{yniguez2014}, findings that support the idea that different mass accretion 
 histories are reflected in the halo's dark matter profile \citep{deason2013}.
Moreover, the radial distribution is also affected by the physics
 considered: \cite{lib2010} showed that at given mass resolution,
 hydrodynamics simulations produced satellites populations more
 concentrated than pure dark matter runs.
The radial distribution is an important aspect nonetheless, because
 its concentration affects the probability of finding planar configurations
 of a given thickness.
We will come back to this in Sec. \ref{Probability for the positions} and
show that we can actually correct the simulation for these differences when
comparing with the observations. 
Given the strong impact of satellite population modelling on their
  spatial distribution, we can already expect that this modelling will also
  affect the properties of the planar configurations found.

\subsection{Finding planes and computing their significances}
\label{Search planes methodology and computation of the significance}

\subsubsection{Satellite plane detection method}
\label{Detection planes methodology}

In order to find three-dimensional structures around the host galaxies we
developed a simple method. 
This method can be applied regardless of the volume or the number of
satellites of the model. 
We compute directly the number of satellites in a plane of a given
thickness. 
We generate a random sample of planes. 
All the planes include the host galaxy and are defined by their normal
vector. 
In order to fill uniformly and homogeneously the volume, 100000 random
planes are generated. 
We fix a thickness $2\Delta$ for each plane. 
Then the distance to the planes of all the satellites of the model are
computed. 
A satellite is define as inside the plane if its distance to the plane is
smaller than $\Delta$. 
We fix the thickness in order to be able to detect the plane of Andromeda,
$2\Delta=40$kpc, i.e. slightly more than 3 time the r.m.s. of satellites
distances to the plane as measured by I13. 
For each plane we obtain the number of satellites included inside a fixed
thickness. 
This simple method can be applied, quickly, to every sample of satellites,
observed or simulated, in the same manner. 

\subsubsection{Computing significance: positions}
\label{Probability for the positions}

\begin{figure*}[ht]
\begin{center}
\includegraphics[scale=0.5]{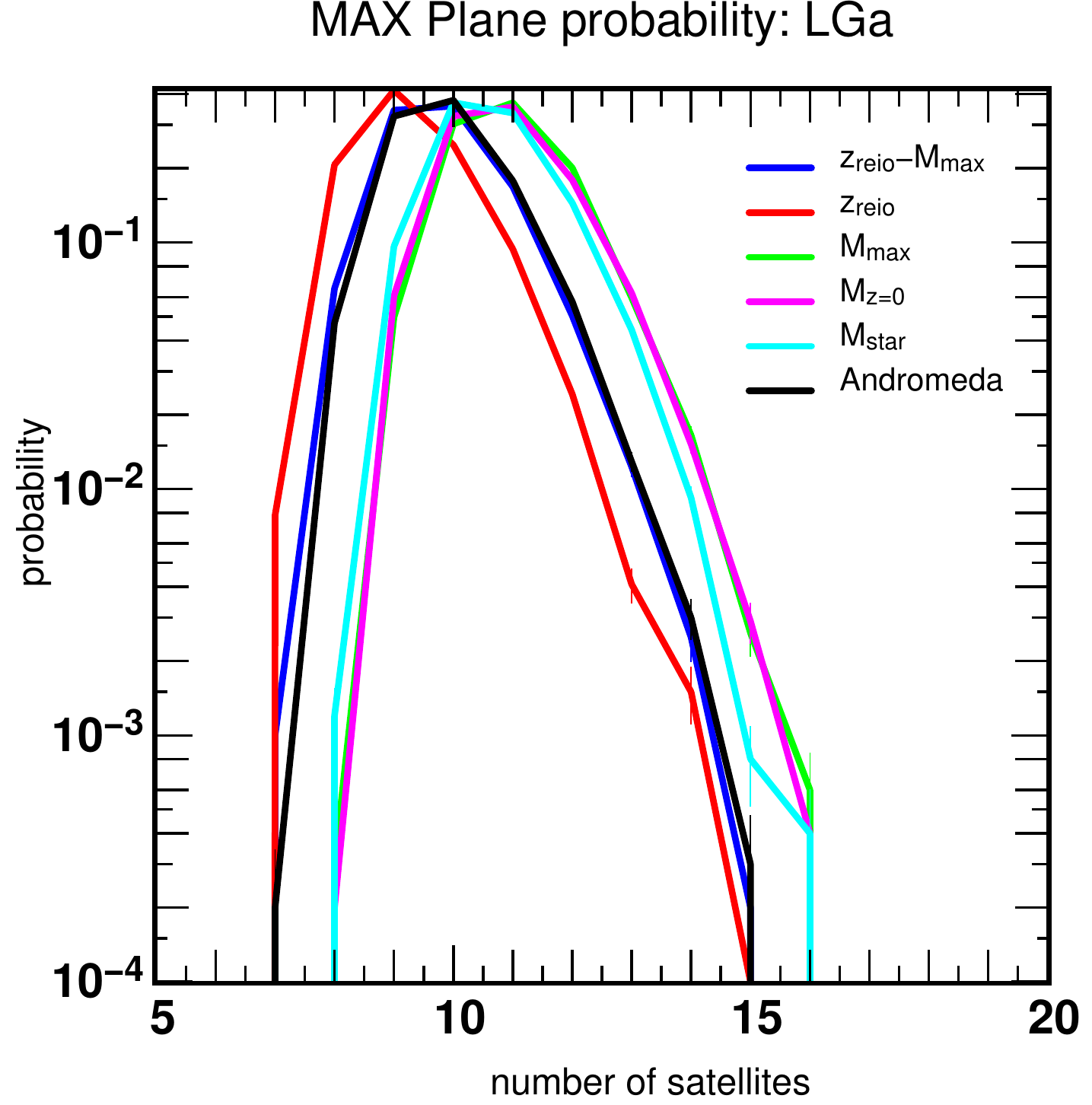}
\includegraphics[scale=0.5]{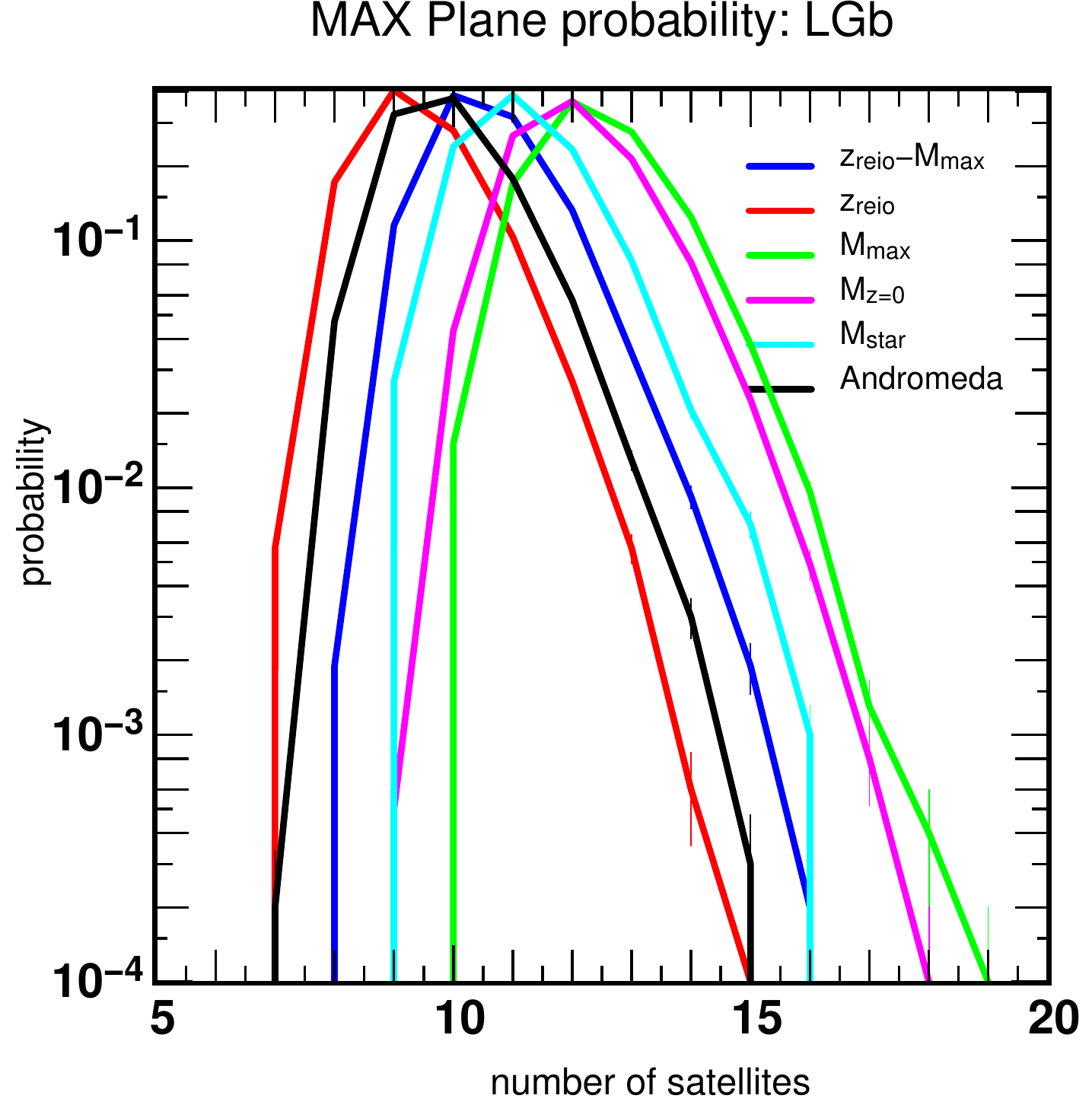}
\caption{\small\label{fig_proba}
The probability distribution of number of satellites in the maximum plane,
for the 10 samples of 25 satellites in a PAndAS volume. 
On the {\em left} panel the 5 samples around LGa, and on the
{\em right}, the 5 around LGb. 
The color code for the different cases tested. 
The black curve on each panel is the probability distribution found for the
observed radial distribution of Andromeda. 
}
\end{center}
\end{figure*}

The plane detection algorithm returns the plane with the largest number of
satellites (which we will refer to as the ``maximum plane''), along with 
the number of satellites it contains $N_{max}$. 
It is tempting to compare directly the simulation's $N_{max}$ with that of
the observed plane. 
In doing this we must however use extreme caution because of one important
caveat: concentrated satellite populations 
tend to naturally have more satellites in any centered plane than extended
satellite populations, 
simply because they are more densely packed. 
Therefore, we also wish to quantify the ``rareness'' of the planar
configurations we find. 
This quantity should allow us to tell whether or not a plane of $N$ satellites
in the simulation is exceptional or not, 
given the radial distribution of the population. 
A good metric of this is the probability of obtaining a similar
configuration in a fully random distribution of satellites.
This is also the metric adopted by I13. 
We will refer to it as the ``significance'' of a plane. 
It is computed for a given detected plane, i.e. for a fixed volume 
(1 of the 3 volumes defined in Sec. \ref{Volume of selection}), and a fixed
satellite population model, 
therefore a fixed radial distribution. 
To compute it we proceed as follows:

We generate randomly $N_{sat}$ satellites respecting the radial distribution
of the model, included in the volume of selection, spherical or PAndAS. 
We apply our maximum plane detection method to this new sample.
This is done 100000 times, each realization producing a different
$N_{max}$. Therefore we obtain a probability distribution function
(hereinafter pdf) of number of satellites $N_{max}$ in the maximum plane for
the fixed radial distribution. 
The Figure \ref{fig_proba} shows the $N_{max}$ probability distributions for
our 5 models applied to both LGa and LGb galaxies, for 25 satellites in a
PAndAS volume. 
In both panels the $N_{max}$ probability distribution of Andromeda (black)
is also shown. 
The shift between the curves is induced by the radial distributions, when
the satellite radial distribution is more concentrated, it is easier to find
planes with a high number of satellites. 
We note that the pdfs are very peaked, with an average number of satellites
in the maximum plane of about 10 satellites. Using these pdfs, we can
compute the $p-value$ of a plane of $k$ satellites as the probability of a
random satellite system to host a plane with $k$ satellites or more. We will
refer to this probability as the positional or spatial $p-value$:

\begin{equation}
p_{pos} = p(X\ge k) = \sum_{k}^{N_{sat}} \mathrm{pdf}
\label{p-value}
\end{equation}

\subsubsection{Computing significance: velocities}
\label{Probability for the velocities}

An aspect making the VPoS of I13 even more striking is the apparent
co-rotation of the plane. We can include this property in our definition of
the significance.
Let us consider a plane containing $N_{p}$ satellites. 
Then we assume that the direction of rotation of a satellites in the plane
is equiprobable. 
Therefore the probability for one satellites to rotate one way or the other
is the same as making head or tail when flipping a coin, which follows a
binomial distribution. 
Therefore the probability to find $k$ satellites rotating in the same
direction is given by a binomial distribution. We will refer to this
probablility as the kinematic or co-rotation $p-value$:

\begin{equation}
p_{kin} = p(X\ge k) = 2 \times\sum_{i=k}^{N_{p}} p(i), 
\label{proba_corot}
\end{equation}

with:

\begin{equation}
p(i) = 
\left(\!\!\!
\begin{array}{c}
N_{p} \\
i
\end{array}
\!\!\!\right) 
\lambda^{i}(1-\lambda)^{N_{p}-i}.
\label{binomial}
\end{equation}

The probability $p_{kin}$, to find $k$ or more co-rotating satellites in a
plane of $N_{p}$ satellites is defined by Equation \ref{proba_corot} and
\ref{binomial}, taking $\lambda =0.5$. 
We multiply by 2 because we do not fix a preferential rotation way, both are
possible.
With these conventions, $p_{kin}(k)$ only has a meaning for $k \ge N_{p}/2$.

\subsubsection{Validation on the observed VPoS of M31}
\label{Planes detection in the observations of M31}

\begin{figure*}[!htbp]
\begin{center}
\includegraphics[scale=0.4]{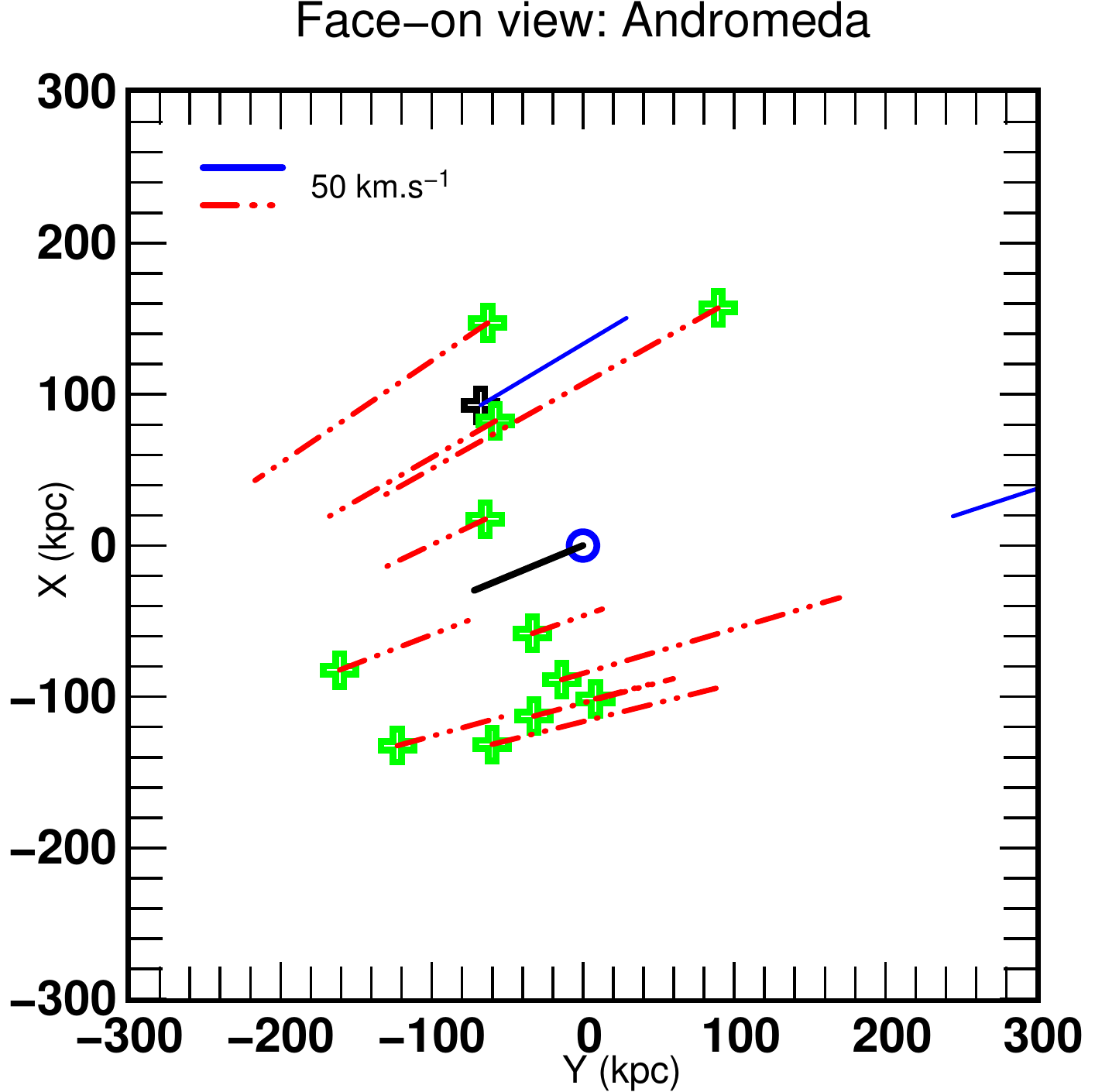}
\includegraphics[scale=0.4]{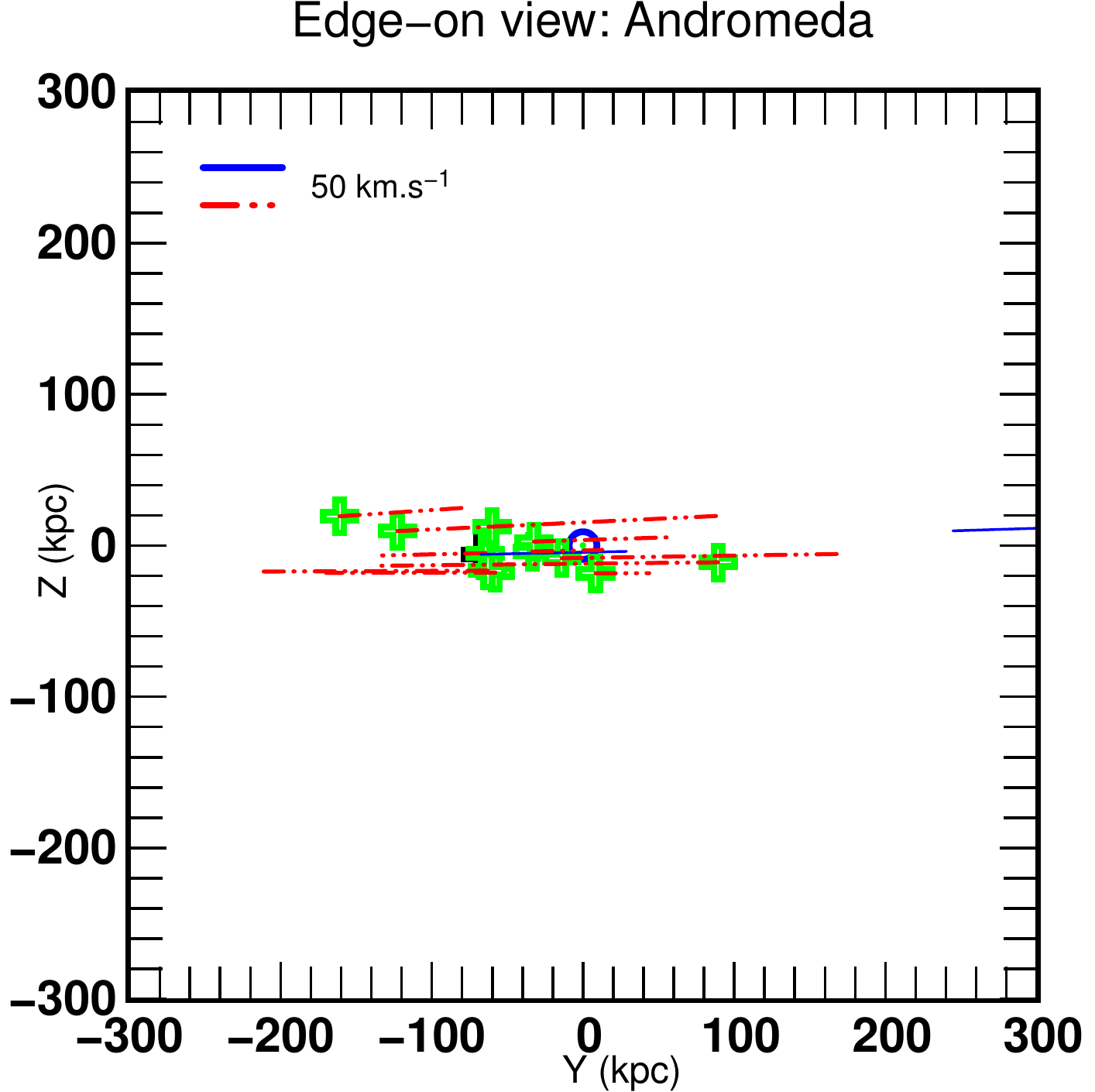} 
\includegraphics[scale=0.4]{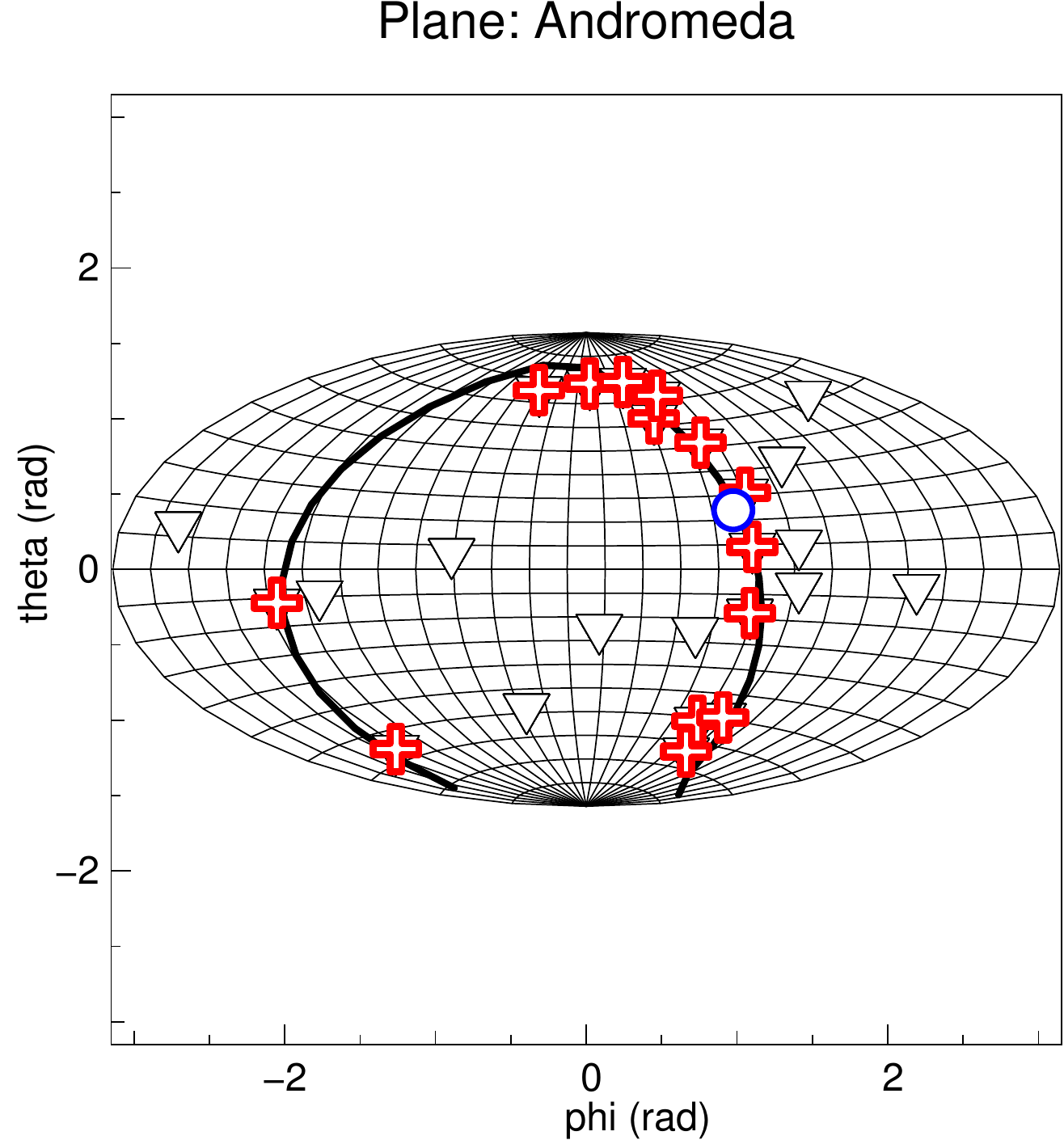}
\caption{\small\label{fig_And}
The observed plane of Andromeda (I13) as detected by our method. 
The face-on and edge-on view of the plane are presented on the {\em top
    left and right} panels. 
Only the satellites of the plane are shown (crosses), along with their
velocities. 
The color of the satellites (green for the dominant rotation vs black) and
their velocities give their rotation 
directions. 
The blue circle with a black line shows the center of M31 and the direction
towards the Milky Way.
{\em Bottom} panel: the satellites of the plane are shown in red in an
Aitoff-Hammer projection showing the positions of M31’s satellites. 
The positions show where each object would appear in the sky if viewed from
the center of M31.
We detect 14 satellites, with 12 co-rotating. 
There are two additional satellites out of the boxes in the {\em top
    left and right} panels, one on the right and an other on the left. 
We recall that only the line of sight velocity is known, and the plane is
seen edge on. 
This alignment of the plane with the direction to the Milky Way (blue
circle) can also be seen in the right panel. 
The properties of this plane are given at the bottom of Table
\ref{tab_compa_25_pandas}. 
}
\end{center}
\end{figure*}

Here we validate our plane detection method by trying to detect the known
plane of Andromeda (I13). 
We compiled the M31 satellites data by taking the (l,b) coordinates from
\cite{McCo2006}, the distances from Tab. 2 of \cite{conn2012}, and the
velocities from Tab. 5 of \cite{col2013}.
Figure \ref{fig_And} shows the maximum plane detected by our method.
We find 14 satellites in a plane of 40kpc of thickness, with 12 co-rotating
satellites. 
Here we do not take into account the distance uncertainties. Because of
this, we do not find exactly the same plane as detected by I13. Indeed I13's
plane hosts one additional satellite, AND III.
This detail left aside, our method reliably recovers detect the existing
plane of satellites of Andromeda.
\par The two last lines of Table \ref{tab_compa_25_pandas} present our
computation of the $p-values$ for this detection of Andromeda's plane, along
with the $p-values$ published by I13.
This configuration of 14 satellites in a plane has a probability to occur of
1.6\% assuming a random distribution. Accounting for the 12 co-rotating
satellites, the total probability is 0.0208\%.  
I13 compute by Monte Carlo the probability for a plane of 15 satellites with
13 co-rotating in a sample of 27 satellites. They find $p-values$ of 0.13\%
for the planar structure and 0.74\% for the co-rotation, which make a total
probability of 0.00096\%. 
The difference we find with respect to I13 is mainly due to AND III, which
is included in I13's plane, but not in ours, due to a slightly different
formulation of plane membership.
If we use the probability density function we computed for the observed
radial distribution and for 15 satellites, we find a probability of 0.33\%
for the planar structure. Multiplied by the probability of co-rotation of 13
satellites (0.75\%), we find a probability for this structure of 0.0024\% to
occur in a random population, which is more compatible with the estimation
of I13. 

 \renewcommand{\tabularxcolumn}[1]{>{\centering\arraybackslash}m{#1}} 
{\setlength{\extrarowheight}{3pt}
\begin{table*}[!htbp] 
\begin{center} 
\begin{tabularx}{\textwidth}{cXXXcXXXXccX} 
\hline 
\hline 
(1) & (2) & (3) & (4) & (5) & (6) & (7) & (8) & (9) & (10) & (11) & (12) \\ 
Galaxy & Model & $N_{max}$ & $N_{cor}$ & RD $\chi^{2}$ & $\Phi$ & $p_{pos}$ (\%) & $p_{kin}$ (\%) & $p_{tot}$ (\%) & $\sigma_{\parallel}$ (kpc) & $\sigma_{\perp}$ (kpc) & $\mathrm{{L_{LOS}}^{max}_{min}}$ \\ 
\hline \\ [-6 pt] 
 & $\mathrm{z_{r}M_{max}}$ & 9 & 5 & 1.15 & 54.3 & 93.75 & 100 & 93.7 & 187.4 & 13.4 & $0.63^{0.99}_{0.1}$ \\ [+3 pt] 
 & $\mathrm{z_{r}}$ & 10 & 7 & 4.40 & 57.7 & 37.8 & 34.3 & 13.0 & 209.5 & 15.5 & $0.62^{0.96}_{0.37}$ \\ [+3 pt] 
 LGa & $\mathrm{M_{max}}$ & 12 & 7 & 1.80 & 15.4 & 28.47 & 77.4 & 22.0 & 104.9 & 11.5 & $0.59^{0.86}_{0.32}$ \\ [+3 pt] 
 & $\mathrm{M_{z=0}}$ & 11 & 7 & 1.79 & 42.6 & 62.42 & 54.8 & 34.2 & 125.3 & 11.2 & $0.5^{0.76}_{0.28}$ \\ [+3 pt] 
 & $\mathrm{M_{star}}$ & 12 & 7 & 1.29 & 62.5 & 20.31 & 77.4 & 15.7 & 160.7 & 13.9 & $0.6^{0.87}_{0.23}$ \\ [+3 pt] 
 \hline \\ [-6 pt] 
 \rowcolor[gray]{0.90}[1.\tabcolsep]  & $\mathrm{z_{r}M_{max}}$ & 11 & 10 & 1.90 & 65.6 & 51.26 & 1.1 & 0.6 & 176.5 & 15.6 & $0.9^{1.55}_{0.56}$ \\ [+3 pt] 
 & $\mathrm{z_{r}}$ & 11 & 8 & 1.39 & 102.1 & 13.91 & 22.6 & 3.1 & 170.1 & 13.8 & $0.56^{0.95}_{0.22}$ \\ [+3 pt] 
 LGb & $\mathrm{M_{max}}$ & 14 & 10 & 4.40 & 111.7 & 16.77 & 17.9 & 3.0 & 126.3 & 11.6 & $0.72^{1.03}_{0.42}$ \\ [+3 pt] 
 & $\mathrm{M_{z=0}}$ & 13 & 8 & 3.96 & 114 & 32.6 & 58.1 & 18.9 & 123 & 11.7 & $0.79^{1.25}_{0.42}$ \\ [+3 pt] 
 & $\mathrm{M_{star}}$ & 11 & 7 & 2.82 & 130.6 & 72.61 & 54.8 & 39.9 & 116.8 & 15.4 & $0.8^{1.2}_{0.19}$ \\ [+3 pt] 
 \hline \\ [-8 pt]
 M31 & observed              & 14 & 12 &      X &  88.5 & 1.60 & 1.3  & 0.0208  & 154.7 & 12.5 & 1.47    \\ 
 M31 & I13               & 15 & 13 &      X &  89   & 0.13 & 0.74 & 0.00096 & 191.9 & 12.6 & 1.3               \\ 
 \hline
\end{tabularx} 
\caption{\label{tab_compa_25_pandas} 
Detected plane within the PAndAS area with $N=25$ satellites. 
The columns (1) and (2) present the host galaxy and the type of selection. 
The columns (3) and (4) shows the detected planes, with receptively the number of satellites in the plane and the number of 
co-rotating satellites. 
The columns (5) presents a qualitative deviation from the radial distribution of the selection to the observations. 
The columns (6) presents the angle between the normal vector of the planes and the line of sight. 
Then the columns (7) and (8) are p-values for the position and the co-rotation. 
The column (9) show the total probability of the detection, including probabilities from position (7) and co-rotation (8). 
Finally columns (10), (11) and (12) present geometrical parameters used as selection in \citet{ibata2014}. 
They present the parallel and perpendicular rms and the minimum specific angular momentum.
Column (12) gives the angular momentum in units of $\times10^4$ km $\mathrm{s^{-1}}$ kpc. 
The two last line are dedicated to the observed plane of Andromeda. The first is for our detection of the plane and our estimation of significance.
The last second is the detection of I13 and their own estimation of p-values. 
} 
\end{center} 
\end{table*}

Having described the simulation, plane detection method and validated the
latter, we move on to searching planar configurations of satellites in the
simulated galaxies.

\section{Results}
\label{Planes detection}

In this section we apply our plane detection method to our model satellite
populations  in the 3 volumes considered.

\subsection{Planes of satellites in the simulation: 25 satellites, PAndAS volume}
\label{Planes detection in the simulation: 25 satellites, PAndAS volume}

\begin{figure*}[!htbp]
\begin{center}
\includegraphics[scale=0.4]{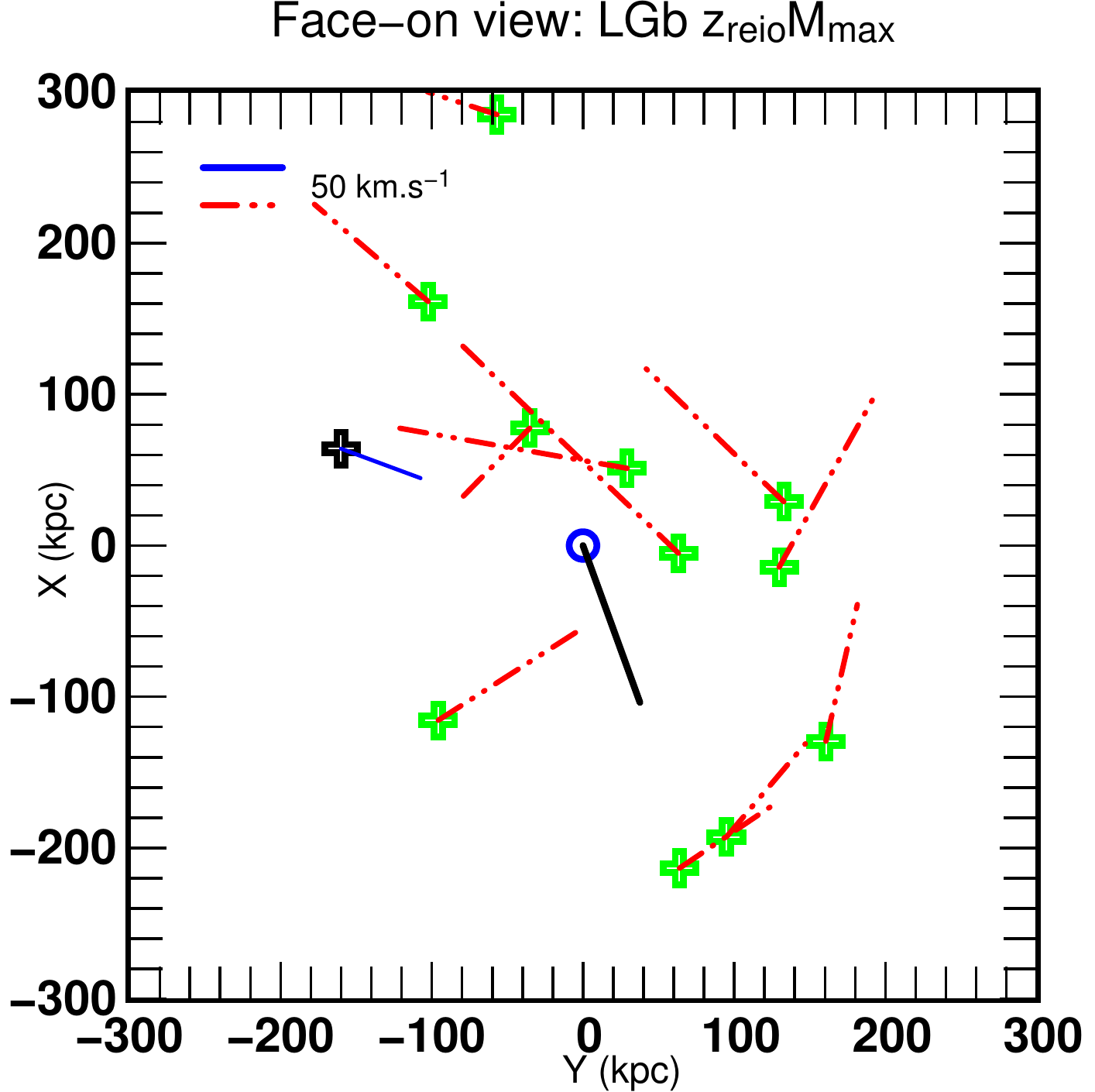}
\includegraphics[scale=0.4]{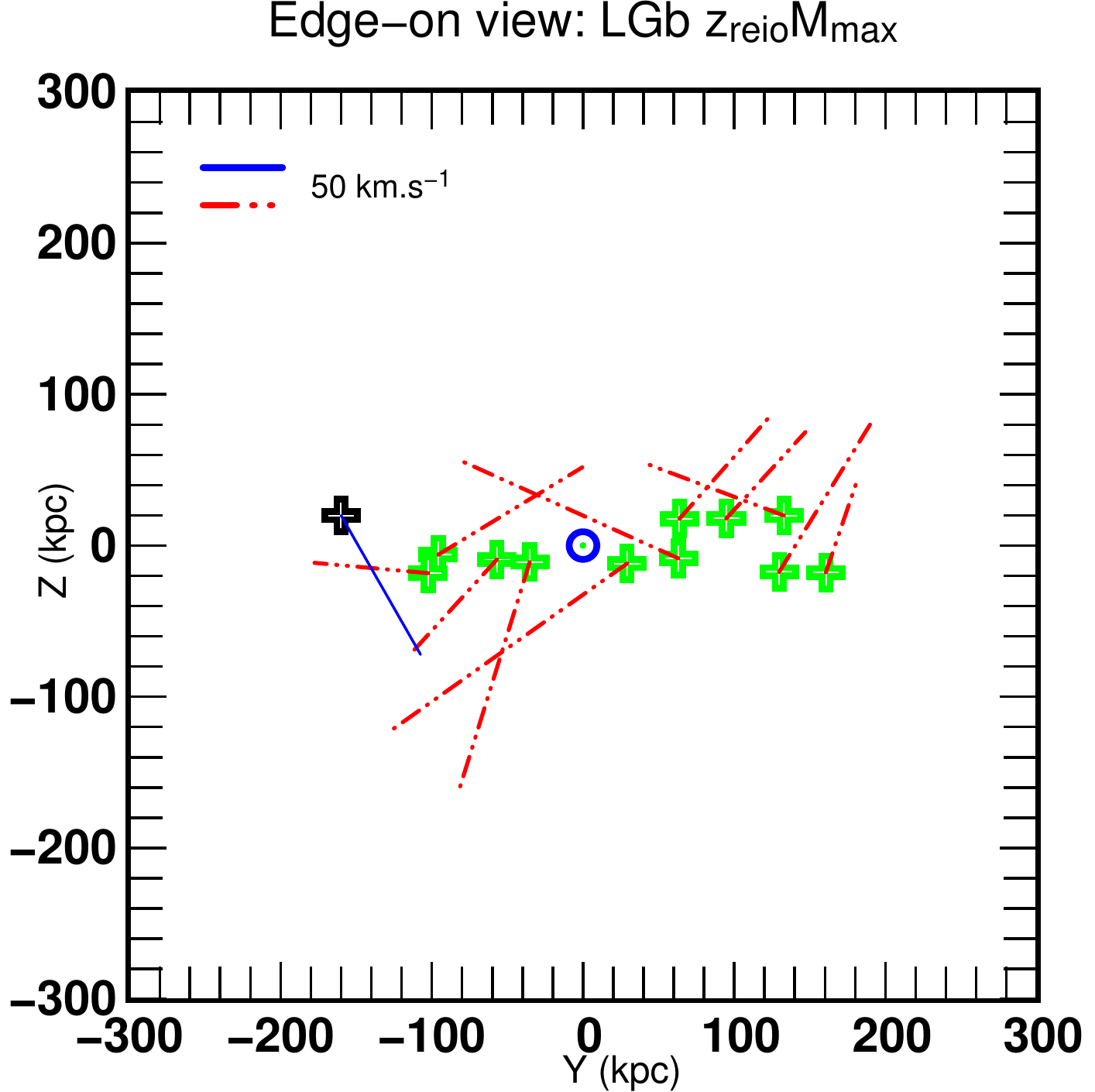} 
\includegraphics[scale=0.4]{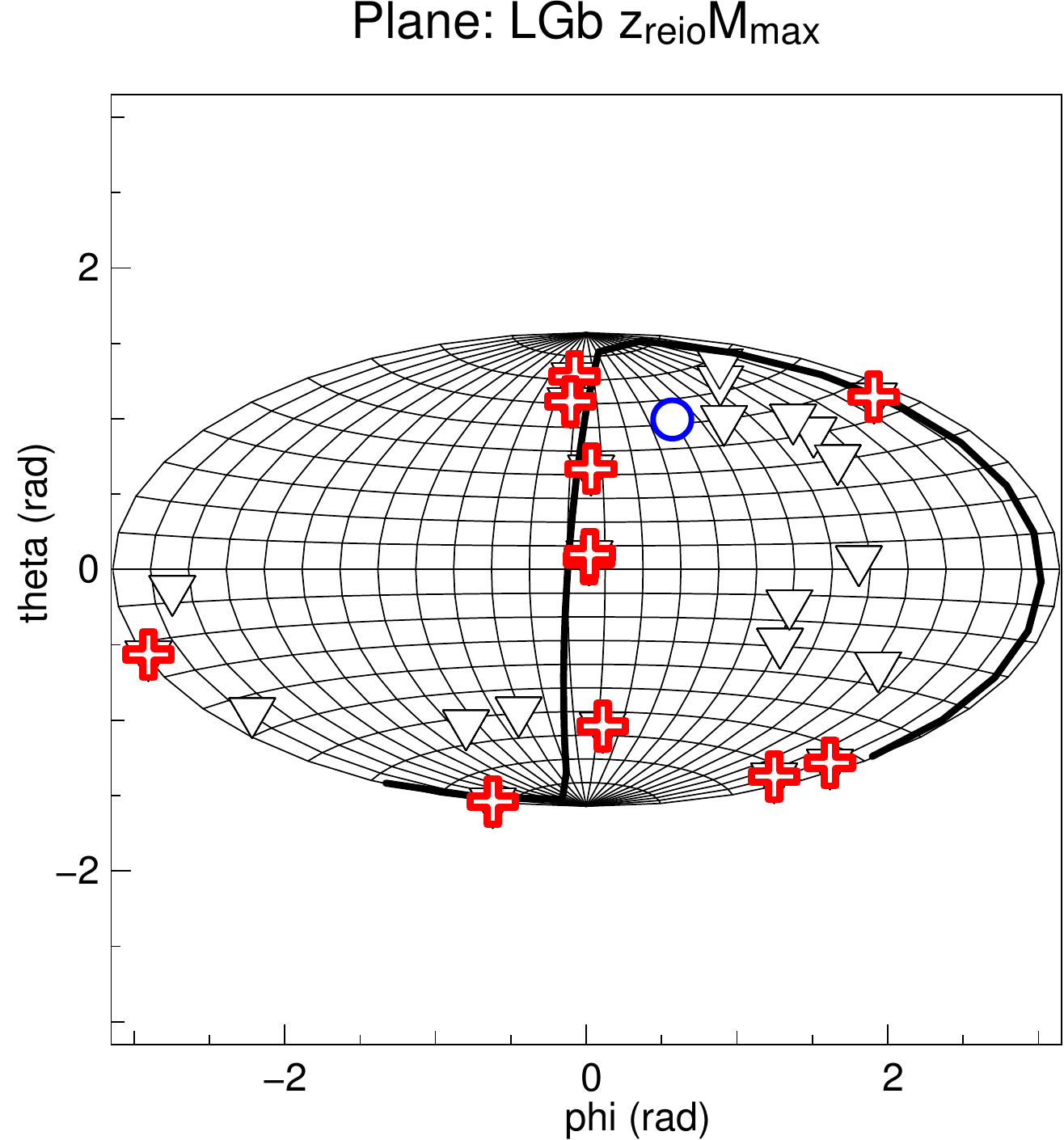}
\caption{\small\label{fig_Ait}
The planes around LGb, detected in the sample of 25 satellites of the ${\rm
  z_{r}}{\rm M_{max}}$ model in the PAndAS volume. 
The face-on and edge-on view of the plane are presented in the {\em top
    left and right} panels. 
Only the satellites of the plane are shown (crosses), along with their
velocities. 
The color of the satellites (green for the dominant rotation vs black) and
their velocities give their rotation directions. 
The blue circle with a black line shows the center of LGb and the direction
towards LGa.
{\em Bottom} panel: the satellites of the plane are plotted in red in
an Aitoff-Hammer projection as viewed from the center of LGb. 
The plane contains 11 satellites, with 10 co-rotating. 
Note that here the velocities are fully known, therefore we can see
velocities pointing away from the plane, while this can not happen with the
observed plane, due to the lack of proper motions.
The orientation of the plane with respect to LGa (blue circle) can also be
seen in the {\em bottom} panel.
Note that the reference frame of the {\em bottom} panel is arbitrary,
but remains fixed in all the figures of the LGb's planes. 
This plane refers to row 6 of Tab. \ref{tab_compa_25_pandas} (${\rm
  z_{r}}{\rm M_{max}}$ model).
}
\end{center}
\end{figure*}

We apply the method for all the satellite population models around the
simulated host galaxies LGa and LGb.
The maximum planes found for LGa and LGb are presented in the Table
\ref{tab_compa_25_pandas}. 
The column (3) of Table \ref{tab_compa_25_pandas} gives $N_{max}$ the number
of satellites found in the maximum plane, of thickness $2\times
\Delta=40$kpc, with a selection of 25 satellites in a PAndAS area. 
\par Firstly, we find strong differences between LGa and LGb. Indeed, the
maximum plane around LGb contains more satellites than  LGa's. We know
already that the two host galaxies have intrinsically different satellite
populations. It can readily be seen in Figure \ref{fig_SDP}, that the radial
density profiles of LGa and LGb are different. 
The distribution of satellites is more extended around LGa. This could be
the cause for its lower number of satellites in the maximum plane.
\par The least populated plane contains 9 satellites, and 14 for the biggest
one. 
It has to be noted that we do not fix the angle between the line of sight
and the detected plane. 
Therefore we can detect planes in any orientation with respect to the line
of sight, and indeed find planes with a variety of orientations, as shown by
column (6) of Table \ref{tab_compa_25_pandas}, which gives  the angle
between the line of sight and the normal vector of the planes. Due to the
axial symmetry, a similar angle does not guarantee that 2 planes have the
same 3d orientation.
\par We do not find planes containing more satellites than the observed
plane of Andromeda, but there is one plane containing 14 satellites. 
We compute the significance of the detections in order to do a proper
comparison (see section \ref{Probability for the positions}). 
In a first step we only consider the probability to find a planar structure
(Table \ref{tab_compa_25_pandas} column (7)) assuming the radial
distributions.
Using only the spatial $p-values$ there are no significant planes\footnote{We
  recall that a $3\sigma$ plane (respectively $5\sigma$) would have a
  $p-value$ of 0.27\% (respectively 0.00003\%) for a gaussian distribution of
  events. In the current paper we will arbitrarily refer to {\em
    significant} planes as having a $p-value$ less than 1\%.}.
The smallest spatial  $p-values$ are for the plane of 14 satellites of the LGb
${\rm M_{max}}$ model (16.77\%) and for the LGb ${\rm z_{r}}$ model
(13.91\%) which contains only 11 satellites. 
\par The effect of the radial distributions is again illustrated by the
spatial $p-values$ of the planes LGb ${\rm z_{r}}{\rm M_{max}}$ and ${\rm
  z_{r}}$: both contain 11 satellites, but the former has a spatial $p-value$
of 51.26\% versus 13.91\% for the latter. 
This means that for the radial distribution of LGb ${\rm z_{r}}$ it is more
difficult to find a structure of 11 satellites than in LGb ${\rm z_{r}}{\rm
  M_{max}}$. And indeed the radial distribution of LGb ${\rm z_{r}}$ is more
dilute. 
We now proceed to include the kinematic properties of the planes in
assessing their significance.

\subsubsection{Velocity in the detected planes}
\label{Velocity in the detected planes}

An important aspect of the plane of Andromeda is the fact that 13 satellites
of the 15 satellites of the plane seem to co-rotate (I13). 
In the observations, only the line of sight velocity is accessible. 
In the simulation, all three components of the velocity are fully known. 
Therefore, it is possible to exactly determinate the number of co-rotating
satellites, unlike the observations.
The column (4) of Table \ref{tab_compa_25_pandas} gives the number of
co-rotating satellites for each detected planes and the $p-value$ for the
co-rotation $p_{cor}$ in column (8) (see Section  \ref{Probability for the
  velocities}). 
The total significance or $p-value$ of a given plane is the product of the 2
other $p-values$, spatial and co-rotation. It is given in column (9) of
Tab. \ref{tab_compa_25_pandas}. It is a more meaningful assessment of the
significance of the planes found.

Here we consider that a total probability lower than 1\% is a significant
detection. Even if it is three decades above the observed planes, this still
means that the detected planar configuration appears in only 1 in 100
realizations of a random satellite population. Therefore, finding such an
alignment purely by chance is still very rare.
\par None of the simulation planes has a total significance as small as the
observations. 
But, there is one case of significant co-rotation. 
Indeed, in one of the planes, 10 satellites of a plane of 11 are rotating
the same way, giving a probability of co-rotation of 1.1\% (gray line in
Table \ref{tab_compa_25_pandas}). 
But the spatial $p-value$ is 51.26\% which give a total probability to occur
at random for this plane of 0.60\%. 
This significant detection is around LGb, for the ${\rm z_{r}}{\rm M_{max}}$
model. 
\par This plane is the example of the fact that the probability of
co-rotation permit, to planes that are not interesting in term of planar
structure, to become significant. 
An other aspect is the sensitivity of the probability of co-rotation to
small variation in the number of objects. 
Indeed 7 satellites co-rotating over 11 give a probability of 54.8\%, while
8 over 11 give 22.6\% and 10 over 11 is 1.1\%. 
A variation of one satellite can change the probability of co-rotation by
more than 30\%. Figure \ref{fig_Ait} shows the face-on and edge-on view of
the most significant plane. The face-on view illustrates the co-rotating
nature of the plane. 

We now explore the geometrical properties of the planes with the
observations.

\subsubsection{Properties of the most significant planes}
\label{Properties of the significant detected planes}

We use tree additional parameters to compare the properties of the planes found
in our simulation with the observations, in the spirit of \cite{ibata2014}:
\begin{itemize}
\item{plane thickness $\sigma_{\perp}$, computed as the perpendicular rms of
  the satellites distance to the plane.}
\item{plane size $\sigma_{\parallel}$, computed as the dispersion of
  galacto-centric distances of the plane satellites.}
\item{$L_{LOS}$: the specific angular momentum for velocities evaluated
  along a line of sight. Therefore the planes are not necessarily seen
  edge-on. To compute comparable values of $L_{LOS}$, an edge-on line of
  sight has to be taken. Once the edge-on line of sight is fixed, we compute
  the median of the product between the velocities projected on the line of
  sight, and the distances in the plane to the host, as describe in
  \citet{ibata2014}. We perform this for 200 random line of sights, and
  retain the average of $L_{LOS}$ on the 200 line of sight, with the maximum
  and minimum values, such as given in column (12) of Table
  \ref{tab_compa_25_pandas}.}
\end{itemize}

We compute these parameters for our detection of the Andromeda's plane, but
we do not take into account ANDXXVII in the computing of the parallel rms
because the error is too large. 
\par We compute these parameters for the maximum plane of all our models
(Table \ref{tab_compa_25_pandas} columns (11), (12) and (13)). 
Our most significant plane contains 11 satellites with 10 that are
co-rotating, giving a probability to occur at random of 0.58\% (Table
\ref{tab_compa_25_pandas}, gray line). 
For this plane, we find $\sigma_{\parallel}=176.5$ kpc,
$\sigma_{\perp}=15.6$ kpc and 
$\mathrm{L_{LOS}}=1.55\times10^4 \mathrm{km.s^{-1}.kpc}$ in the most
favorable case. These values compare rather well with the observed plane.

\subsubsection{Conclusion for 25 satellites in PAndAS volume}
\label{Conclusion for 25 satellites in PAndAS volume}

In this first exploration of the planes of satellites in our simulation,
considering five different models for the satellite population and using a
pseudo-survey volume as close as possible to PAndAS, we find 1 rather
exceptional plane, with a total probability to occur in a random population
lower than 1\%.
This finding, in a simulation with only two major disk galaxies, suggests that
the satellite population is not random and anisotropic, but highly
structured.
This plane is geometrically comparable to the observed plane in thickness
and size. 
However, it contains only 11 satellites, 10 co-rotating, and therefore has a
statistical significance (quantified by the total $p-value$) lower than the
observed plane of I13. The simulation, while successfully reproducing some
degree of structure in the satellite populations, does not yield satellite
planes as extreme as the observed VPoS of M31.

In the rest of the paper, we will allow ourselves to modify the volumes and
the number of satellites considered in order to analyse further the
structure of the satellite populations of the simulation.

\section{Planes of satellites in alternative volumes}
\label{Looking for planes in alternative volume selections}

The definition of the PAndAS volume imposes a line of sight and angular
limits. 
\par In this section we will explore the structure of the satellite
population in a slightly extended PAndAS volume, which we will refer to as
PAndAS-bis.
The inner and outer limits of the cone are 2.5 deg (2 deg in the PAndAS
volume), and 12 deg (10 deg in the PAndAS volume).
We also change the MW-M31 distance, setting it to 1200 kpc, which is the
distance between the two galaxies in the simulation.
The resulting volume is larger than the PAndAS volume. Indeed, the projected
inner and outer limits are now 42 kpc and 207 kpc.
Because the PAndAS-bis volume is larger, we will consider samples of 27
satellites, which is the number of satellites detected by I13 in the real
PAndAS volume, which include an additional area around M33 compared to the
quasi-circular area centered on M31.
Therefore this volume allows us to investigate what could be found in
slightly more remote satellite populations.
\par In a second step we will leave aside PAndAS and PAndAS-bis conical
volumes and consider a simple spherical volume around the host. The sample
consists of the satellites found in the 50-500 kpc shell.
Finally, we will also explore more abundant satellite populations, by
setting $N_{sat}=$25, 27, 30, 35, 50, 100 and 150 satellites. Indeed, since
the faintest MW satellites are still about 100 times fainter than the
faintest M31 satellites known, one can only expect that future deeper
surveys will discover new, fainter satellites hiding in the PAnDAS
area. Therefore we take advantage of the current study and simulation to
investigate the degree of structure this new population may display.

\subsection{PAndAS-bis volume with 27 satellites}
\label{PAndAS-2 volume with 27 satellites}

\renewcommand{\tabularxcolumn}[1]{>{\centering\arraybackslash}m{#1}} 
{\setlength{\extrarowheight}{3pt}
\begin{table*}[!htbp] 
\begin{center} 
\begin{tabularx}{\textwidth}{cXXXcXXXXccX} 
  \hline
  \hline
  (1) & (2) & (3) & (4) & (5) & (6) & (7) & (8) & (9) & (10) & (11) & (12)\\ 
  Galaxy & Model & $N_{max}$ & $N_{cor}$ & RD $\chi^{2}$ & $\Phi$ & $p_{pos}$ (\%) & $p_{kin}$ (\%) & $p_{tot}$ (\%) & $\sigma_{\parallel}$ (kpc) & $\sigma_{\perp}$ (kpc) & $\mathrm{{L_{LOS}}^{max}_{min}}$ \\ 
  \hline \\ [-6 pt] 
   & $\mathrm{z_{r}M_{max}}$ &  9 &  6 & 2.66 & 125.3 & 84.2 & 25.4 & 21.4   & 222.7 & 12.0 & $0.62^{1.0 }_{0.09}$ \\ [+3 pt]
  
   & $\mathrm{z_{r}}$        &  9 &  6 & 7.88 & 126.3 & 47.3 & 25.4 & 12.0   & 266.1 & 13.3 & $0.77^{1.08}_{0.18}$ \\ [+3 pt]
  
  LGa & $\mathrm{M_{max}}$      & 10 &  7 & 0.54 & 129.0 & 90.2 & 17.2 & 15.5   & 125.0 & 14.6 & $0.60^{0.90}_{0.15}$ \\ [+3 pt]
  
   & $\mathrm{M_{z=0}}$         & 10 &  6 & 1.10 & 136.7 & 82.0 & 37.7 & 30.9   & 178.3 & 11.7 & $0.79^{1.19}_{0.35}$ \\ [+3 pt]
  
   & $\mathrm{M_{star}}$        & 10 &  8 & 1.12 & 104.5 & 88.9 & 5.5  &  4.9   & 184.7 & 15.3 & $0.54^{0.72}_{0.26}$ \\ [+3 pt]
  \hline \\ [-6 pt]
  \rowcolor[gray]{0.90}[1.\tabcolsep] & $\mathrm{z_{r}M_{max}}$ & 14 & 11 & 1.11 & 55.0  & 0.55 & 2.9  & 0.016 & 176.3 & 14.1 & $1.23^{1.98}_{0.42}$ \\ [+3 pt]
  
  \rowcolor[gray]{0.90}[1.\tabcolsep] & $\mathrm{z_{r}}$        & 10 & 10 & 4.49 & 90.5  & 24.4 & 0.1  & 0.024 & 212.7 & 14.0 & $0.98^{1.49}_{0.55}$ \\ [+3 pt]
  
  \rowcolor[gray]{0.95}[1.\tabcolsep] LGb & $\mathrm{M_{max}}$      & 13 & 11 & 1.02 &  66.2 & 33.0 & 1.1  & 0.36 & 140.4 & 12.7 & $0.72^{0.99}_{0.40}$ \\ [+3 pt]
  
  \rowcolor[gray]{0.95}[1.\tabcolsep] & $\mathrm{M_{z=0}}$         & 14 &  9 & 0.71 &  69.6 & 3.96 & 13.3 & 0.53 & 145.1 & 13.7 & $0.80^{1.38}_{0.40}$ \\ [+3 pt]
  
   & $\mathrm{M_{star}}$        & 12 &  7 & 0.76 &  49.6 & 45.8 & 38.7 & 17.7   & 123.8 & 15.4 & $0.76^{1.11}_{0.32}$ \\ [+3 pt]
  \hline
\end{tabularx}
\caption{\label{tab_PAndAS2}
Same as Tab. \ref{tab_compa_25_pandas} for 27 satellites in the PAndAS-bis
volume.
}
\end{center}
\end{table*}

\begin{figure*}[!htbp]
\begin{center}
\includegraphics[scale=0.4]{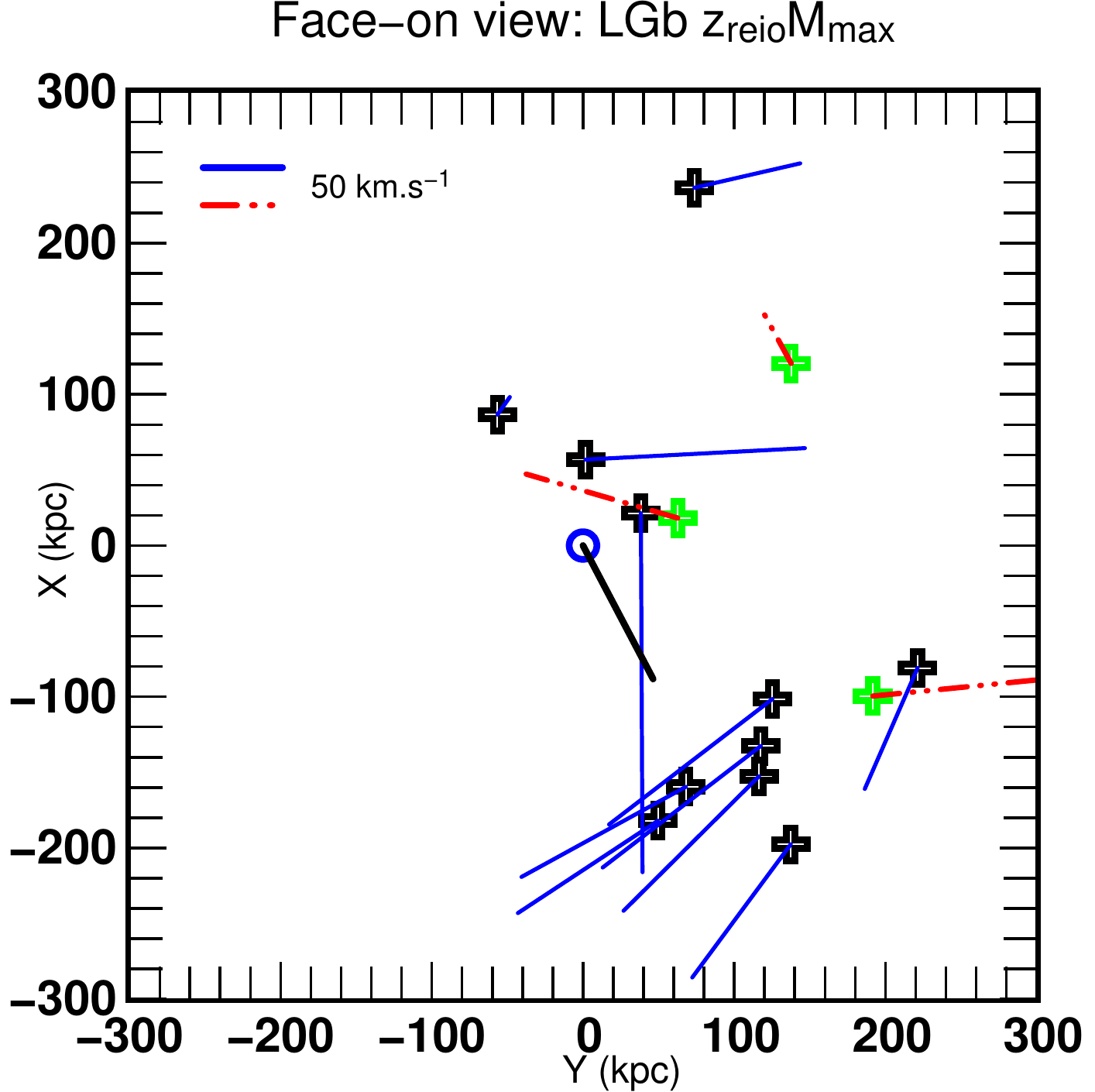}
\includegraphics[scale=0.4]{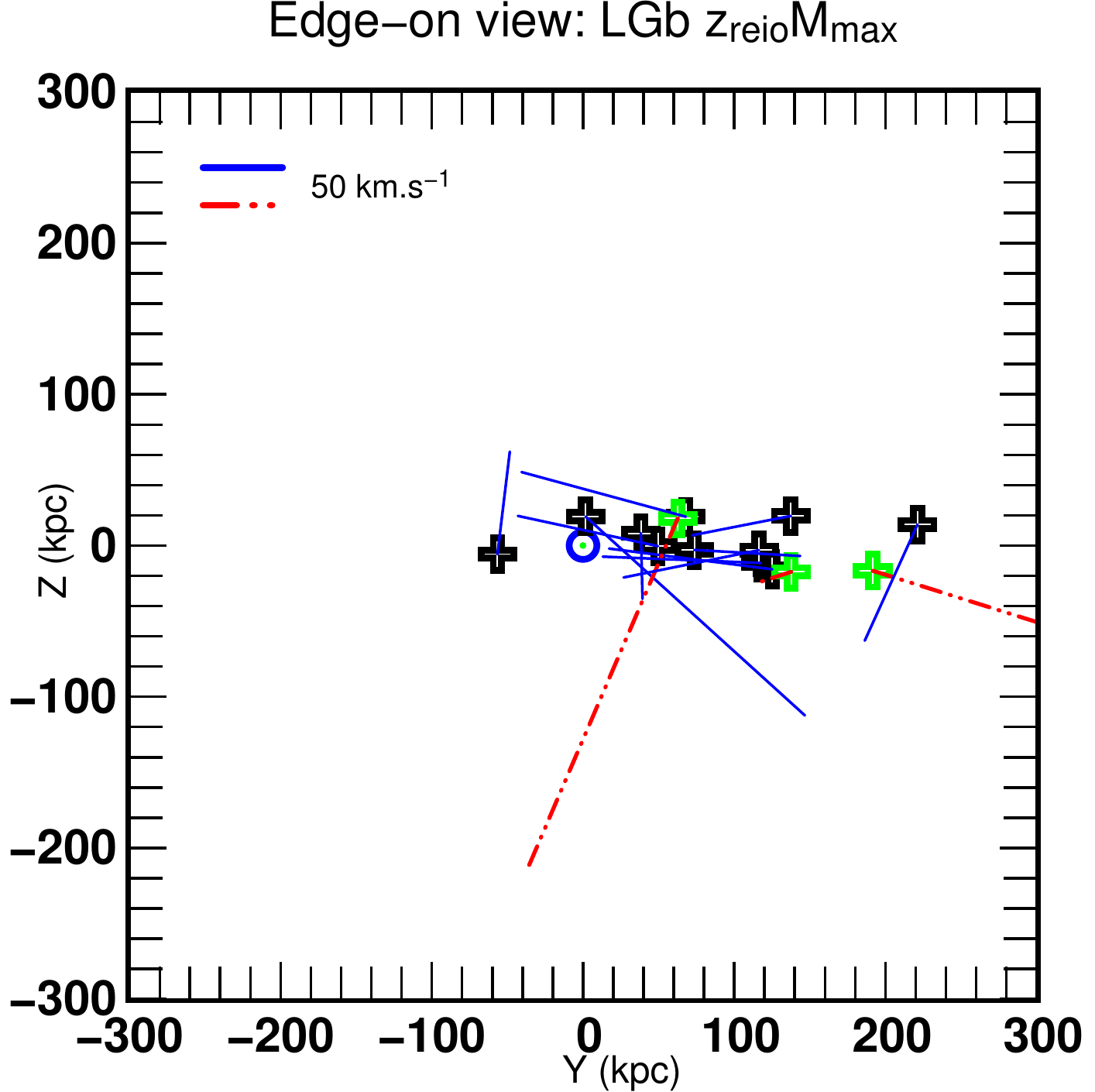} 
\includegraphics[scale=0.4]{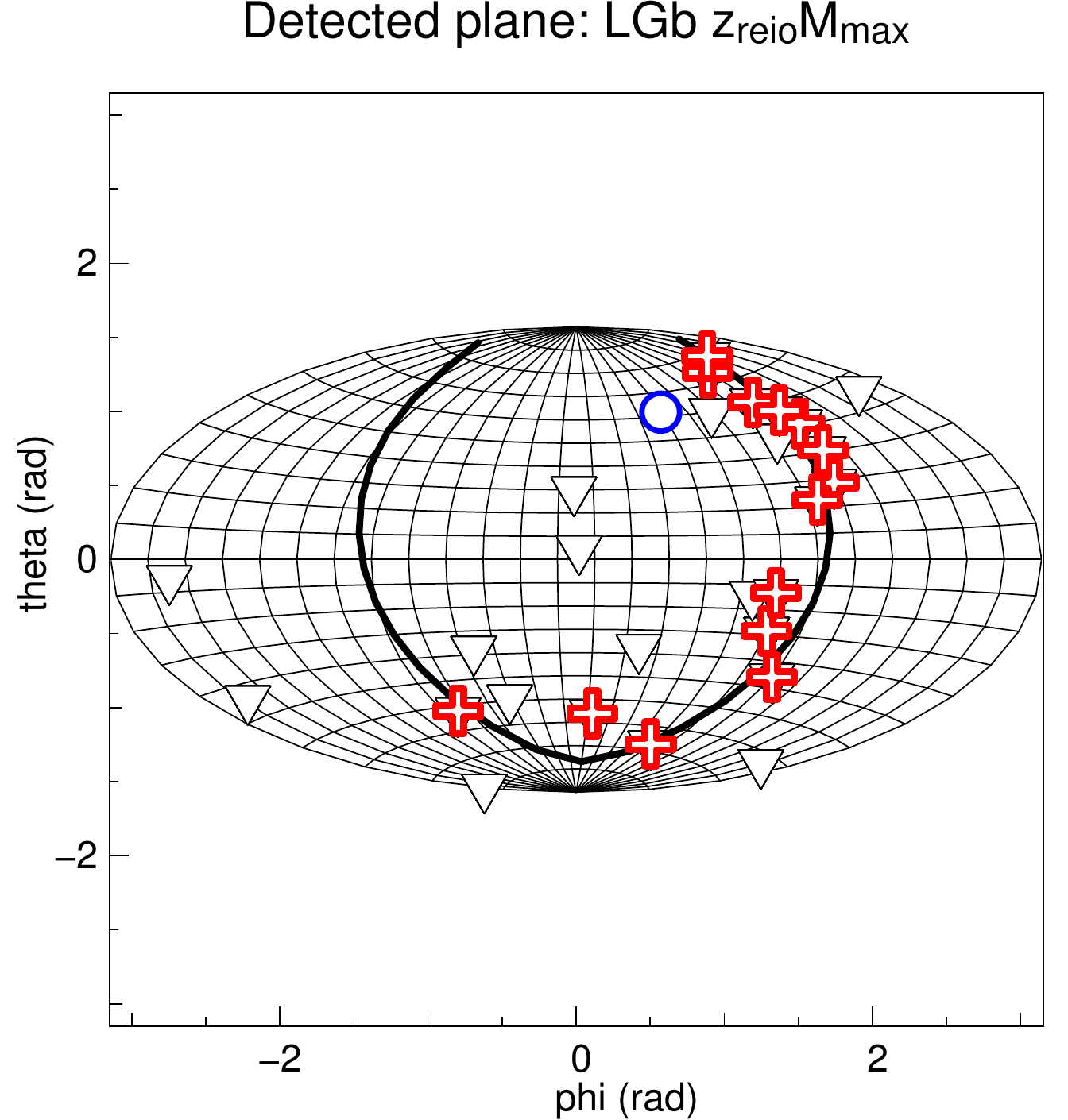}
\caption{\small\label{fig_ZM_PAndAS2}
The plane of satellites around LGb, detected in the sample of 27 satellites
of the ${\rm z_{r}}{\rm M_{max}}$ selection in the PAndAS-bis volume. 
The face-on and edge-on view of the plane are presented on the {\em top
    left and right} panels. 
Only the satellites of the plane are represented by crosses and segments
indicating the in-plane component of the velocities. 
The color of the symbols code their direction of rotation, which can also be
judged from the velocity vectors. 
{\em Bottom} panel: the satellites of the plane are plotted in red in
an Aitoff-Hammer projection as viewed from the center of LGb.
We detect 14 satellites, with 11 co-rotating. 
The {\em bottom} panel also shows the orientation of the plane with
respect to LGa (blue circle). 
Note that the reference frame of the {\em bottom} panel is arbitrary,
but remains fixed in all the figures of the LGb's planes. 
This figure refers to the plane described in line 6 of Table
\ref{tab_PAndAS2}.}
\end{center}
\end{figure*}

\begin{figure*}[!htbp]
\begin{center}
\includegraphics[scale=0.4]{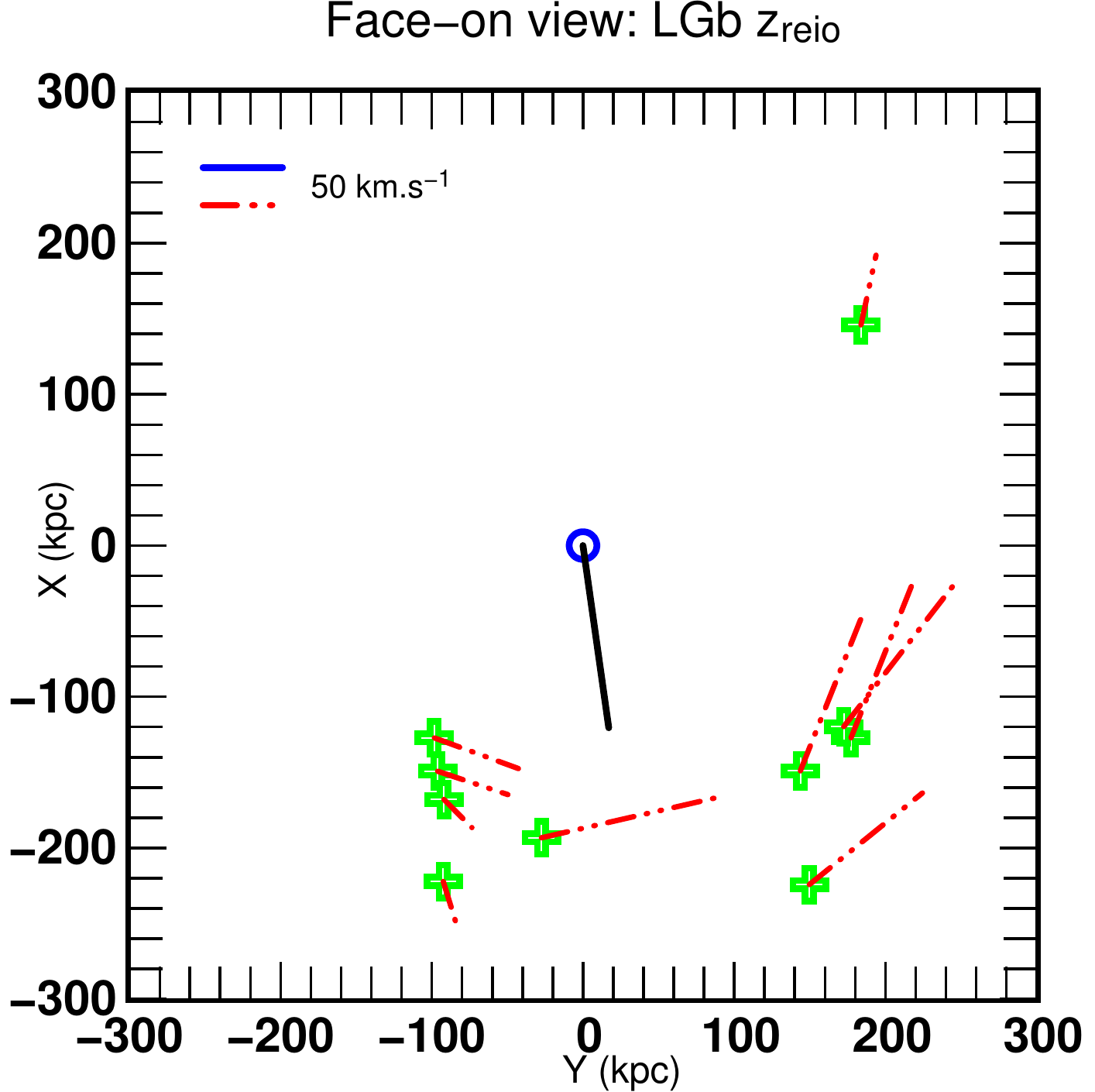}
\includegraphics[scale=0.4]{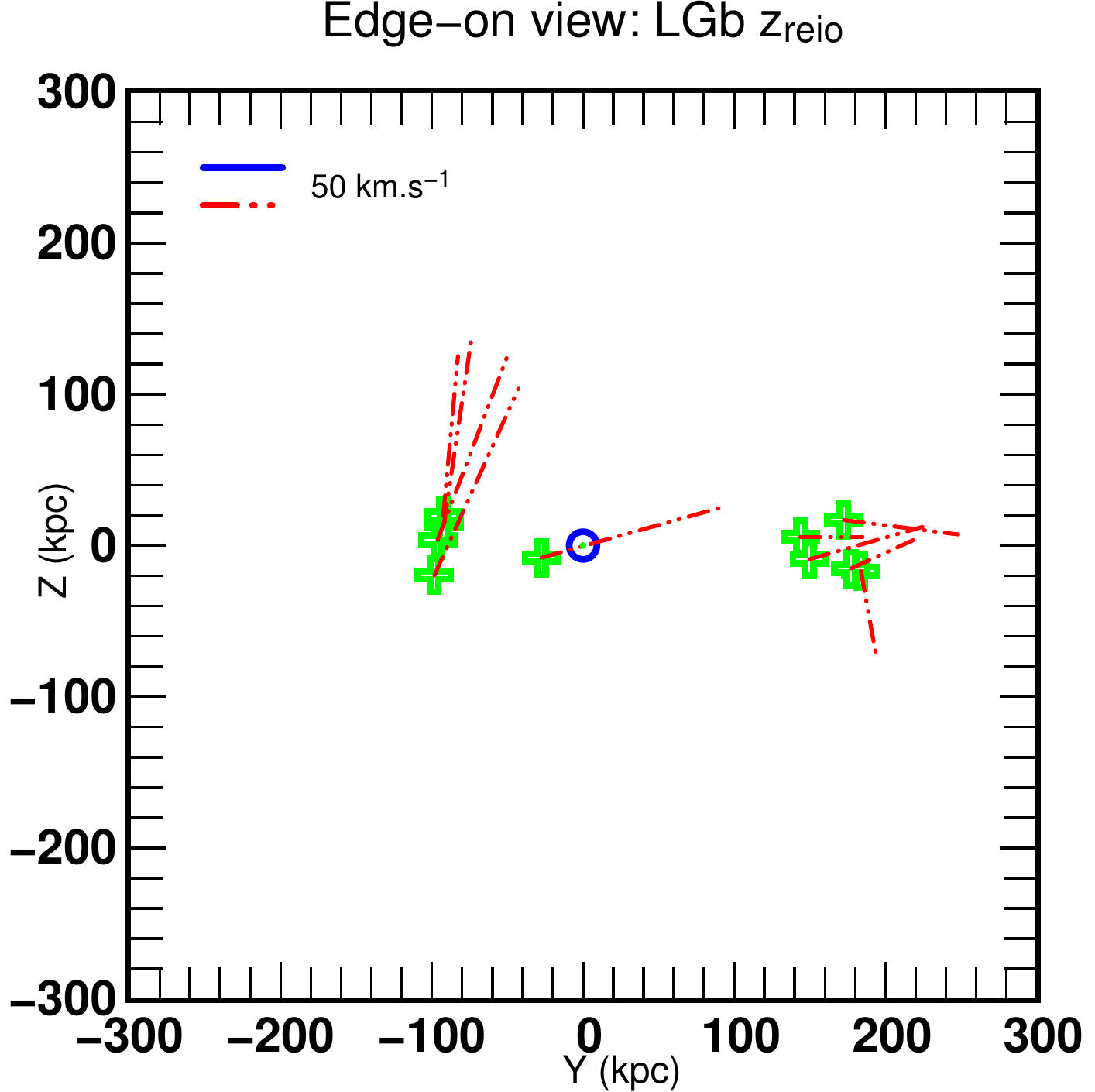} 
\includegraphics[scale=0.4]{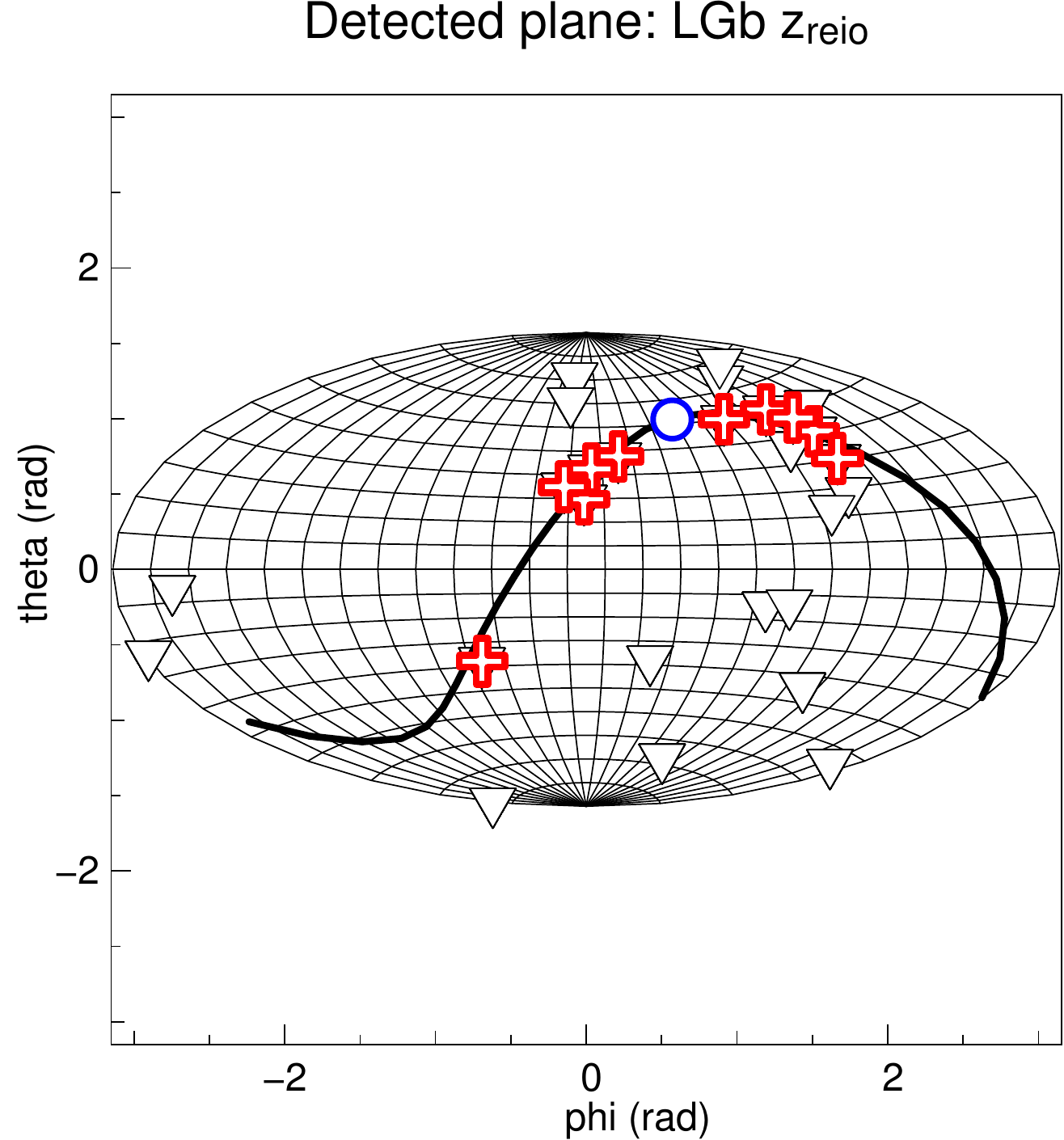}
\caption{\small\label{fig_Z_PAndAS2}
The planes around LGb detected in the sample of 27 satellites of the ${\rm
  z_{r}}$ model in the PAndAS-bis volume. 
The face-on and edge-on view of the plane are presented in the {\em top
    left and right} panels. 
Only the satellites of the plane are shown (crosses), along with their
velocities. 
The color of the satellites (green for the dominant rotation vs black) and
their velocities give their rotation directions. 
The blue circle with a black line shows the center of LGb and the direction
towards LGa.
{\em Bottom} panel: the satellites of the plane are plotted in red in
an Aitoff-Hammer projection as viewed from the center of LGb. 
The plane contains 10 satellites, with 10 co-rotating. 
The {\em bottom} panel also shows the orientation of the plane with
respect to LGa (blue circle).
Note that the reference frame of the {\em bottom} panel is arbitrary,
but remains fixed in all the figures of the LGb's planes. 
This plane refers to row 7 of Tab. \ref{tab_PAndAS2} (${\rm z_{r}}$ model).}
\end{center}
\end{figure*}

The Table \ref{tab_PAndAS2} gives the properties of the planes detected in
the PAndAS-bis volume with samples of 27 satellites. 
\par Firstly, as in Section \ref{Planes detection in the simulation: 25
  satellites, PAndAS volume}, the two galaxies of the simulation give very
different results.
The detected planes are more abundant around LGb, with 10 to 14 satellites
in the planes, compare to 9 to 10 in LGa. This reflects an intrinsical
difference in the distributions of the satellites around the two galaxies:
LGa's population is more extended spatially, making rich planes of a given
thickness more rare.
\par We find four planes with probabilities to occur at random lower than
1\%, from which two are lower than 0.1\% (gray and dark-gray lines in Table
\ref{tab_PAndAS2}). 
We will now analyse the two most significant planes of LGb, corresponding to
models ${\rm z_{r}}M_{Max}$ and ${\rm z_{r}}$. 
The LGb ${\rm z_{r}}M_{Max}$ plane contains 14 satellites with 11
co-rotating, while the LGb ${\rm z_{r}}$ plane only contains 10 satellites
but all of them are co-rotating. 
These two planes are interesting because they are highly significant for two
quite different reasons. 
The first has a high structural significance ($p_{pos}$), while the second
stands out due to its kinematic properties (100\% co-rotation).
\par Both planes seem to made of groups of satellites.
This is illustrated by the top left panel of Figure \ref{fig_ZM_PAndAS2},
featuring one large group of six satellites (around y$=100$ kpc, x$=-100$
kpc). 
The top right panel shows that the velocities of the group are mostly within
the plane. There are however five satellites with strong out of the plane
velocities, which will therefore get out of the plane on a short timescale.
Therefore the nature of this plane is dual: it is mainly composed of one
coherent group travelling together, plus a handful of additional satellites
which are accidentally aligned with it at the time of observation or
analysis.
This apparent clumpiness of the satellites in the plane is not unlike the
observed VPoS of M31: this can be seen by comparing the top left panel of
Figure \ref{fig_And} and \ref{fig_ZM_PAndAS2}, at least for the x$<0$ half
of the figure.
The velocities on the other hand can not be compared so bluntly since only
the line of sight velocities are available in the observations.
The structure of the other plane (LGb ${\rm z_{r}}$) is shown in Figure
\ref{fig_Z_PAndAS2}: it consists of two groups of satellites, plus two
accidentally aligned satellites.
On the top left and right panel, one group has in-plane velocities, while
the other group's velocities point perpendicular to the plane. 
These tow groups are internally coherent, but are otherwise unrelated.
They form the backbone of the plane, the two additional satellites belonging
to the plane are pure chance.
The thick black line of the left panel of Fig. \ref{fig_ZM_PAndAS2} shows
the direction to LGa. From this panel we can infer that the LGb ${\rm
  z_{r}}$ plane would display line-of-sight kinematics qualitatively similar
to the observed VPoS if observed from LGa (i.e. receding on one side and
approaching on the other side), yet it is clearly not a disc and the plane
has no kinematical coherence as a whole.

\par In both cases, the planes are almost aligned with the other galaxy LGa,
as shown by the bottom panel of Figures \ref{fig_ZM_PAndAS2} and
\ref{fig_Z_PAndAS2}. 
We also note that the satellites of the planes are not symmetrically
distributed. 
Indeed most of the satellites are located in the near half rather than the
far half with respect to the other galaxy, as in the observed VPoS. However
we restrain from interpreting this since we have only two clear occurences at
hand.

As a preliminary conclusion, we see that the simulated galaxies feature
planes of satellites which have some degree of similarity in richness and
geometry with the observed VPoS of M31, although their $p-values$ are not as
low. These planes consist of one or more coherent satellite groups, although
the groups themselves are unrelated. Therefore they are not coherent discs,
even if they can appear as such if only line of sight velocities are
available.

\subsection{Spherical volume}
\label{Spherical volume}

 \renewcommand{\tabularxcolumn}[1]{>{\centering\arraybackslash}m{#1}} 
{\setlength{\extrarowheight}{3pt}
\begin{table*}[hbt] 
\begin{center} 
\begin{tabularx}{\textwidth}{cXXXcXXXXccX} 
\hline 
\hline 
(1) & (2) & (3) & (4) & (5) & (6) & (7) & (8) & (9) & (10) & (11) & (12) \\ 
  Galaxy & Model & $N_{max}$ & $N_{cor}$ & RD $\chi^{2}$ & $\Phi$ & $p_{pos}$ (\%) & $p_{kin}$ (\%) & $p_{tot}$ (\%) & $\sigma_{\parallel}$ (kpc) & $\sigma_{\perp}$ (kpc) & $\mathrm{{L_{LOS}}^{max}_{min}}$ \\ 
\hline \\ [-6 pt] 
\rowcolor[gray]{0.90}[1.\tabcolsep]  & $\mathrm{z_{r}M_{max}}$ & 32 & 21 & 1.01 & 63.4 & 0.3 & 11 & 0.033 & 203.7 & 13.4 & $0.69^{0.97}_{0.31}$ \\ [+3 pt] 
\rowcolor[gray]{0.90}[1.\tabcolsep]  & $\mathrm{z_{r}}$ & 32 & 21 & 1.25 & 63.4 & 0.07 & 11 & 0.0077 & 212.6 & 13.3 & $0.74^{1.01}_{0.31}$ \\ [+3 pt] 
 LGa & $\mathrm{M_{max}}$ & 32 & 20 & 0.61 & 52.2 & 7.1 & 21.5 & 1.5 & 138.5 & 12.8 & $0.72^{0.86}_{0.51}$ \\ [+3 pt] 
 & $\mathrm{M_{z=0}}$ & 30 & 20 & 0.58 & 69.3 & 20.65 & 9.8 & 2.0 & 152.6 & 13.1 & $0.57^{0.75}_{0.43}$ \\ [+3 pt] 
 & $\mathrm{M_{star}}$ & 27 & 16 & 0.63 & 52.1 & 93.41 & 44.2 & 41.3 & 174.2 & 11.8 & $0.75^{1.1}_{0.37}$ \\ [+3 pt] 
 \hline \\ [-6 pt] 
  & $\mathrm{z_{r}M_{max}}$ & 28 & 19 & 1.52 & 118.1 & 27.96 & 8.7 & 2.4 & 140.2 & 13.8 & $0.93^{1.26}_{0.6}$ \\ [+3 pt] 
 & $\mathrm{z_{r}}$ & 28 & 19 & 1.53 & 118 & 21.55 & 8.7 & 1.9 & 140.2 & 13.7 & $0.93^{1.26}_{0.6}$ \\ [+3 pt] 
 LGb & $\mathrm{M_{max}}$ & 33 & 19 & 1.63 & 130.6 & 9.09 & 48.6 & 4.4 & 129.7 & 13.1 & $0.74^{0.99}_{0.39}$ \\ [+3 pt] 
 & $\mathrm{M_{z=0}}$ & 33 & 17 & 1.80 & 130.9 & 9.99 & 100 & 10.0 & 126.7 & 12.6 & $0.7^{0.9}_{0.46}$ \\ [+3 pt] 
 & $\mathrm{M_{star}}$ & 30 & 20 & 1.38 & 101.4 & 18.79 & 9.8 & 1.9 & 168 & 11.9 & $0.8^{1}_{0.63}$ \\ [+3 pt] 
\hline 
\end{tabularx} 
\caption{\label{tab_compa_100_sph} Results for 100 satellites in a spherical volume around the host.
} 
\end{center} 
\end{table*}

\begin{figure*}[!htbp]
\begin{center}
\includegraphics[scale=0.5]{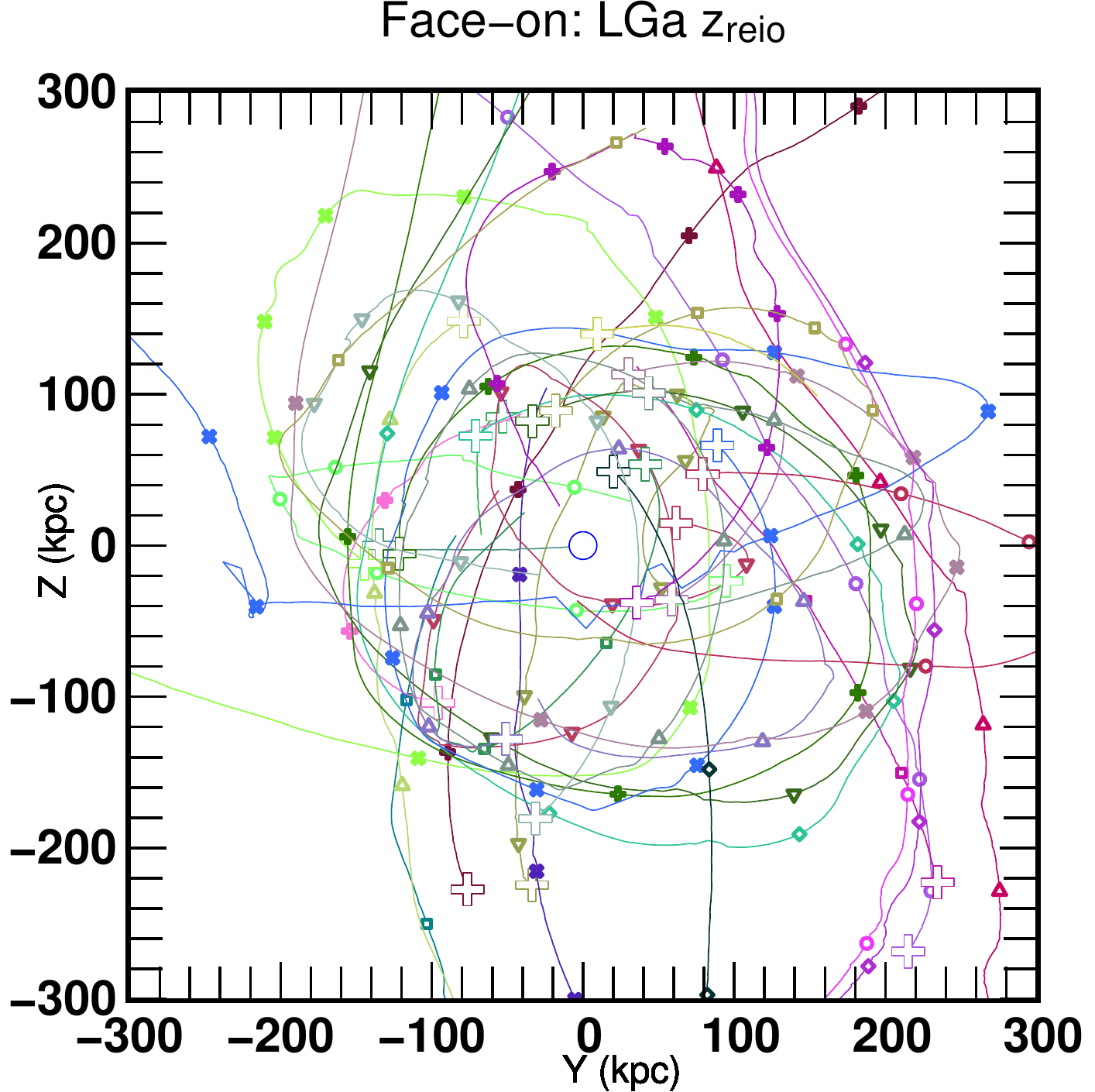}
\includegraphics[scale=0.5]{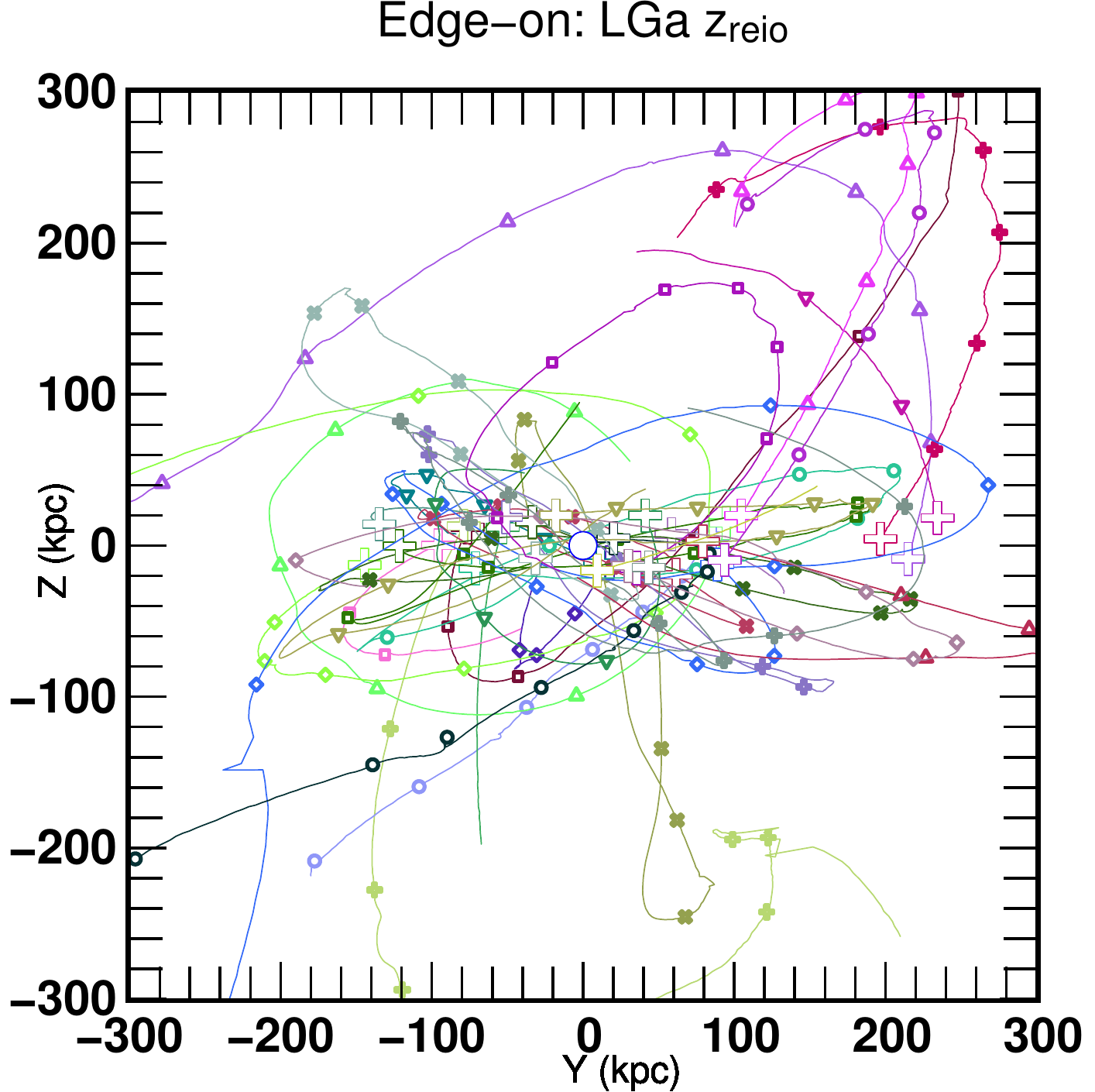} \\
\includegraphics[scale=0.5]{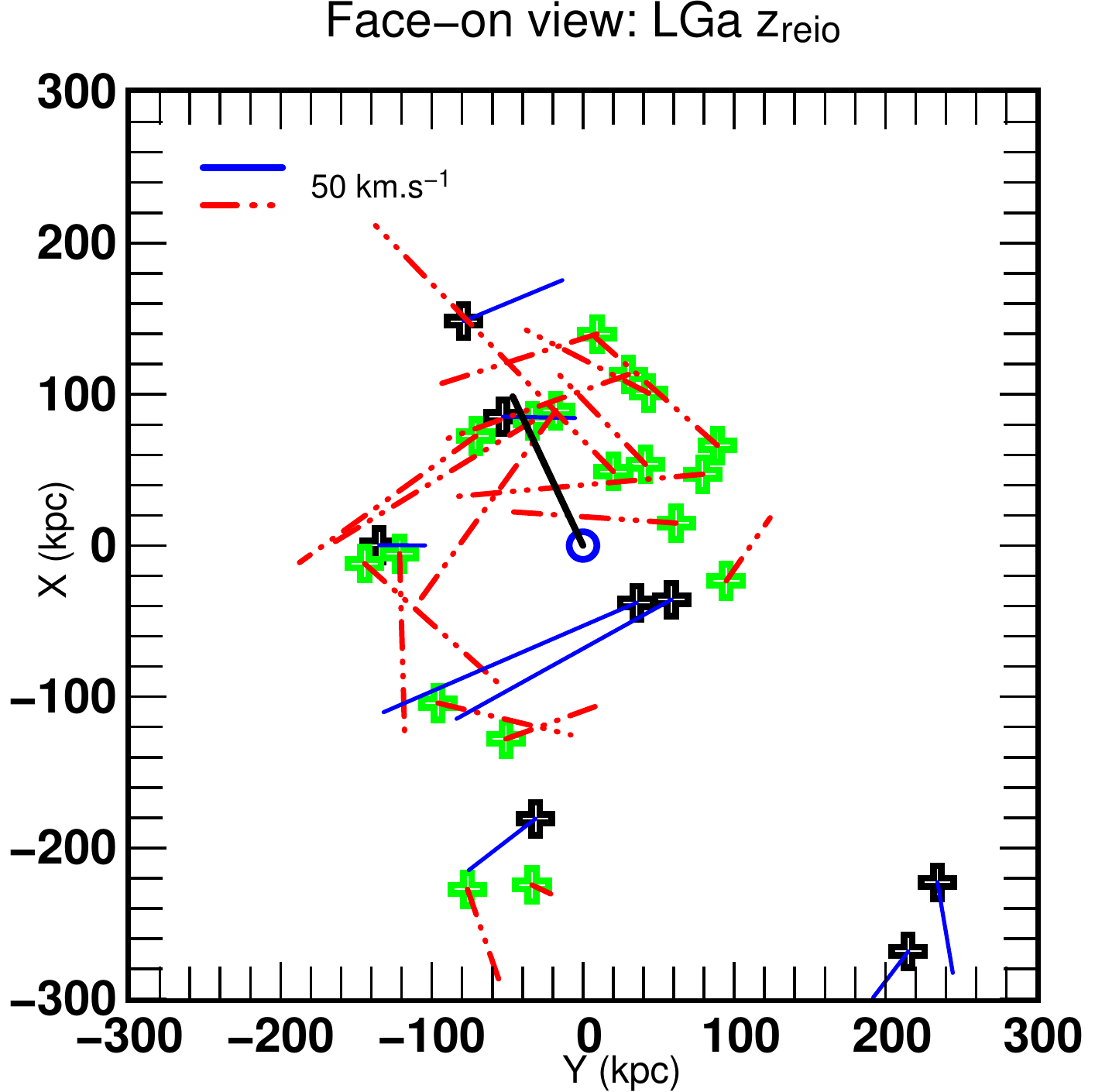}
\includegraphics[scale=0.5]{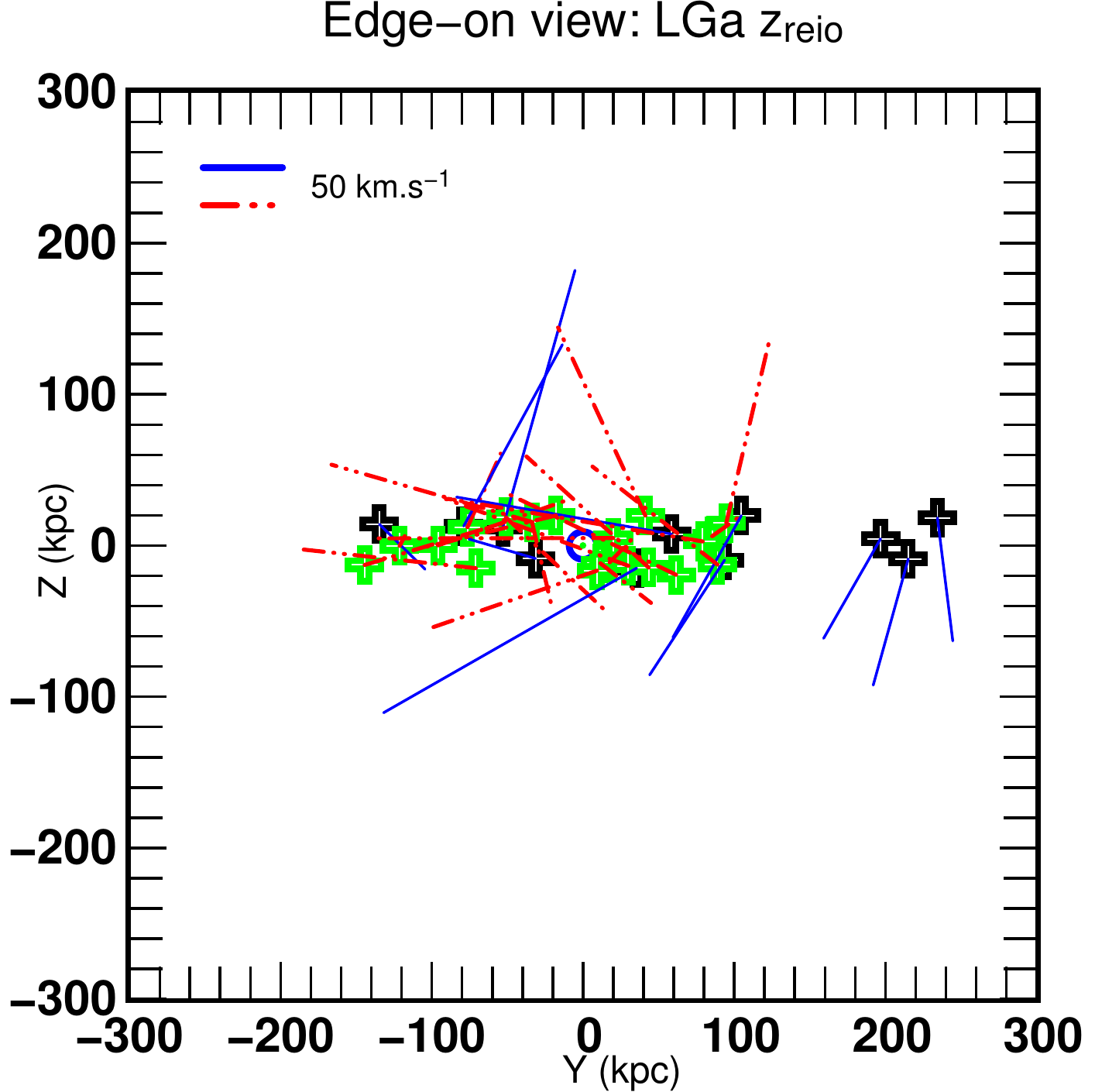} 
\caption{\small\label{fig_100_Z}
The significant planes for the LGa ${\rm z_{r}}$ model for $N_{sat}=$ 100
satellites. 
The {\em top row} shows the trajectories seen face-on ({\em left})
and edge-on ({\em right}). 
The cross are the position at {\em z}=0, and the dots on the trajectories are
separated by a 50Myr duration. 
The small hickups along some trajectories are due to problematic halo
identifications during rapid evolution, such as galaxy mergers. 
They do not affect our results.
The {\em bottom row} shows the plane's satellites positions and
velocities at {\em z}=0. 
The color of the satellites and their velocity vectors denote the direction
of rotation. This plane refer to the Table \ref{tab_compa_100_sph} lines 2. }
\end{center}
\end{figure*}

In this section we consider a spherical volume selection in order to explore
the simulation in quest of planar structure, without the possible bias due
to the conical shape of the PAndAS and PAndAS-bis volumes.
We will vary the sample size from 25 to 150.
\par It is important to note that the samples are not independent.
For example, the selection ${\rm z_{r}}$ and ${\rm z_{r}}{\rm M_{max}}$ have
a fair number of satellites in common in the 100 and 150 satellites samples. 
\par In the 70 samples considered ($2\times7\times5$) we find 12 planes with
a probability to occur at random lower than 1\%.
The properties of the planes are presented in Table \ref{tab_signi}.
Due to the finite size of the halo population, two different models can
yield the same maximum plane. Indeed we find that among the 12 planes found,
only seven of them are unique. 
We present the results for the case of 100 satellites in a spherical volume
in the Table \ref{tab_compa_100_sph}. 
In this case one significant maximum plane is detected twice (gray lines in
Table \ref{tab_compa_100_sph}), as indicated by the very similar properties
of both lines and viusal inspection.
However not all the satellites are the same in these two models, which is
why small differences in the geometrical properties (columns (10), (11) and
(12)) are found. 
This plane has a strong structural significance, indeed, 32 satellites over
100, one third, are detected in a plane of 40 kpc of thickness. 
The probability to occur at random is 0.07\% assuming the radial
distribution of the LGb ${\rm z_{r}}$ model. 
Its kinematical coherence is rather weak, with 21 satellites over 32, two
thirds, that are rotating the same way, giving a probability $p_{kin}=$11\%. 
This plane has, finally, a total probability to occur at random of 0.0077\%. 
Therefore, even if its $p-value$ is not as low as the observed VPoS of M31,
the structure and kinematics of the simulated satellites population is
highly non-random.
The plane is geometrically comparable to the observations with a
$\sigma_{\parallel}$ of 212 kpc and a $\sigma_{\perp}$ of 13.3 kpc. 
But it has a small specific angular momentum, with on
average$\mathrm{L_{LOS}}=0.75\times10^4$km.$\mathrm{s^{-1}}$.kpc and a
maximum of only $1.01\times10^4$km.$\mathrm{s^{-1}}$.kpc.
\par The Figures \ref{fig_100_Z} show the face-on and edge-on view of this
maximum plane, with the trajectories of the satellites in the top row, and
the velocities at {\em z}=0 in the bottom row. 
Figure \ref{fig_100_Z} relates to the Table \ref{tab_compa_100_sph}, line 2
(LGa model ${\rm z_{r}}$).
On the bottom left panel, visual inspection reveals a handful of small
coherent group of two to tree satellites. 
On the bottom right, we estimate that a third of the 32 satellites of the
planes have velocities pointing away from the plane, while the remaining 2 
thirds are well aligned.

The bottom left panel shows that the plane is very lumpy, with one large
coherent group of satellites forming the bulk of the plane.

As in the previous section, this figure also shows that the plane would
appear as roughly disc-like if only the line of sight velocities were
considered.(i.e. receding on one side of the host, and approaching on the
other side).

\renewcommand{\tabularxcolumn}[1]{>{\centering\arraybackslash}m{#1}} 
{\setlength{\extrarowheight}{3pt}
\begin{table*}[!htbp] 
\begin{center} 
\begin{tabularx}{\textwidth}{XXXXcXXXXccX} 
  \hline
  \hline
(1) & (2) & (3) & (4) & (5) & (6) & (7) & (8) & (9) & (10) & (11) & (12) \\ 
  Galaxy & Model & $N_{max}$ & $N_{cor}$ & RD $\chi^{2}$ & $\Phi$ & $p_{pos}$ (\%) & $p_{kin}$ (\%) & $p_{tot}$ (\%) & $\sigma_{\parallel}$ (kpc) & $\sigma_{\perp}$ (kpc) & $\mathrm{{L_{LOS}}^{max}_{min}}$ \\ 
  \hline \\ [-6 pt] 
  25-LGb & $\mathrm{z_{r}M_{max}}$ & 12 & 11 & 1.63 & 65.6 & 6.57 & 0.6 & 0.042 & 173.5 & 15.9 & $0.89^{1.33}_{0.5}$ \\ [+3 pt]
  25-LGb & $\mathrm{M_{z=0}}$ & 14 & 9 & 2.57 & 105 & 1.6 & 42.3 & 0.68 & 133.9 & 13.3 & $0.7^{0.97}_{0.31}$ \\ [+3 pt] 
  \hline \\ [-6 pt]
  27-LGb & $\mathrm{z_{r}M_{max}}$ & 12 & 11 & 1.62 & 65.1 & 11.74 & 0.6 & 0.075 & 173.6 & 15.3 & $0.89^{1.33}_{0.49}$ \\ [+3 pt] 
  27-LGb & $\mathrm{z_{r}}$ & 12 & 9 & 1.44 & 99.6 & 3.51 & 14.5 & 0.51 & 170.7 & 15 & $0.76^{1.11}_{0.45}$ \\ [+3 pt] 
  27-LGb & $\mathrm{M_{z=0}}$ & 15 & 10 & 2.67 & 104.8 & 1.27 & 30.1 & 0.38 & 129.9 & 13.8 & $0.62^{0.94}_{0.27}$ \\ [+3 pt] 
  \hline \\ [-6 pt]
  50-LGa & $\mathrm{z_{r}}$ & 15 & 12 & 3.03 & 115.5 & 20.82 & 3.5 & 0.73 & 266 & 14.1 & $0.8^{1.14}_{0.43}$ \\ [+3 pt] 
  \hline \\ [-6 pt]
  100-LGa & $\mathrm{z_{r}M_{max}}$ & 32 & 21 & 1.00 & 63.4 & 0.3 & 11 & 0.033 & 203.7 & 13.4 & $0.69^{0.97}_{0.31}$ \\ [+3 pt] 
  100-LGa & $\mathrm{z_{r}}$ & 32 & 21 & 1.25 & 63.4 & 0.07 & 11 & 0.0077 & 212.6 & 13.3 & $0.74^{1.01}_{0.31}$ \\ [+3 pt] 
  \hline \\ [-6 pt]
  150-LGa & $\mathrm{z_{r}M_{max}}$ & 44 & 26 & 0.61 & 63.3 & 0.57 & 29.1 & 0.17 & 183.2 & 12.7 & $0.66^{0.93}_{0.41}$ \\ [+3 pt] 
  150-LGa & $\mathrm{z_{r}}$ & 42 & 26 & 0.60 & 63 & 1.68 & 16.4 & 0.28 & 186.3 & 12.6 & $0.67^{0.96}_{0.41}$ \\ [+3 pt] 
  150-LGa & $\mathrm{M_{z=0}}$ & 41 & 29 & 0.50 & 70 & 12.64 & 1.1 & 0.15 & 157.9 & 12.3 & $0.61^{0.78}_{0.45}$ \\ [+3 pt] 
  150-LGb & $\mathrm{M_{star}}$ & 43 & 29 & 1.52 & 102.4 & 4.2 & 3.1 & 0.13 & 158.7 & 11.5 & $0.84^{1.07}_{0.55}$ \\ [+3 pt] 
  \hline
\end{tabularx}
\caption{\label{tab_signi}
Significant detected planes within a spherical volume with $N_{sat}$ going
from 25 to 150. 
The columns (1) present the $N_{sat}$ with host galaxy and the (2) present
the type of selection. 
The columns (3) and (4) shows the detected planes, with respectively
the number of satellites in the plane and the number of co-rotating
satellites. 
The columns (5) presents a qualitative deviation from the radial
distribution of the selection to the observations. 
The columns (6) presents the angle between the normal vector of the planes
and the line of sight. 
Then the columns (7) and (8) are $p-values$ for the position and the
co-rotation. 
The column (9) show the total probability of the detection, including
probabilities from position (7) and co-rotation (8). 
Finally columns (10), (11) and (12) present geometrical parameters used as
selection in \citet{ibata2014}. 
They present the parallel and perpendicular rms and the minimum specific
angular momentum. 
}
\end{center}
\end{table*}

\section{Discussion}

\subsection{To disc or not to disc?}
The planes we find in the simulation do not reproduce exactly the observed
properties of the VPoS but in a few cases they come quite close. 
The ${\rm z_{r}}{\rm M_{max}}$ model for the PAndAS-bis selection for
instance, is close to matching the observed VPoS, both in richness,
structure and kinematics, as shown by Tab. \ref{tab_PAndAS2}. 
We recall that in this simulation, the LGb galaxy which hosts the most
realistic planes has a mass of only $7.81 \times 10^{11} M_{\odot}$, while
the bracket for M31's mass goes up to $M_{300}=1.0-1.8 \times 10^{12}
M_{\odot}$ \citep{watkins2010,marel2012, diaz2014}, i.e. possibly two times larger than our
simulated M31. 
It is very likely that this difference in mass affects our results: a more
massive LGb would have a richer, more extended satellite system and larger
angular momentum, potentially bringing the simulation in better agreement
with the observations with respect to VPoSs. We plan to follow up on this by
studying a new CLUES simulation in a forthcoming paper.
Moreover, the fact that we find rich planes of satellites in the current
simulation where only two relevant satellites systems can be studied (LGa's
and LGb's) suggests that such structures are likely to be common. 
It is not clear why our small volume simulation features satellite
  planes not too dissimilar to M31's, whereas previous studies such as
  \cite{ibata2014} showed that they are rare. 
It could be due to numerical resolution, since our simulation is 15 times
better resolved in mass than the Millenium-II. 
Do high resolution simulations produce more significant planes of satellites
than low resolution runs? 
This remains to be shown. 
It could also be pure luck, but it is difficult to compare our results with
those of \cite{ibata2014} since the number of satellites in the plane are
different. 
It could also be an environmental effect, considering paired galaxies
instead of isolated as in \cite{ibata2014}. 
This is not our favored explanation since \cite{pawlow2014b} did not
find this parameter to have a strong impact on satellite distributions.
Our constrained simulation also captures by construction the environment 
of the LG on a larger scale, for instance with the proximity of a galaxy cluster. 
\cite{jaime2011} proposed that the environment may affect the mass assembly history 
of LG galaxies. Could it also affect the frequency of planar satellite configurations? 
To answer this will require comparative studies such as the present work on constrained 
versus unconstrained simulations of galaxy pairs.
In several previous studies the VPoS of M31 is described as a rotating disc
of satellites \citep{conn2013,bowden2013,pawlow2014}. 
One must recall, though, that only the line-of-sight velocities are
currently available for these objects. 
Therefore the rotational support of the plane is unproven, and until proper
motions are available, can not be firmly assessed. 
Provided that the planes of satellites in our simulation and in M31 are of
the same essence, our results provide interesting insight into this
question. 
Indeed, the simulated planes are not coherent kinematical structures. 
They consist mostly a of a group of satellites forming the backbone of the
plane (about half to 2/3 of the plane's satellites), which aligns by chance
with a number of random satellites or in one case with another small group. 
While the main group is indeed a coherent structure resembling a group
accretion event, and moves within the plane, the other chance-aligned
satellites will fly out of the plane in a short timescale: about 150 Myr assuming average perpendicular
velocities of about 100 km/s and a plane thickness of 15 kpc.
Therefore the current plane appears {\em short}-lived, but if its backbone (the main group) is
long-lived, then we expect new satellites to randomly enter the plane while
some others move out. 
In this respect, the plane may still be {\em long}-lived (supported by its main satellite
group), but 1/3 to 1/2 of its satellites are non-permanent members.
Therefore we consider the plane is half-real (i.e. coherent) and half
random, and this is the main result of this work.
They nevertheless display velocity patterns characteristic of rotation if
seen edge-on and only the line-of-sight velocities are considered, with the
opposite sides respectively receding and approaching. 
But they do not qualify as ``discs''.
We are well aware that our simulated planes of satellites are not perfect
matches to M31's, but if they are of similar nature, then our results
suggest that the observed VPoS of I13 is not a disc, and that 1/3 to 1/2 of
its members should have large proper motions perpendicular to the plane (up
to 200 km/s). 
These high proper motion satellites are likely to be the most spatially and
kinematically isolated ones, for instance: 
AndI, AndIII, AndIX, And XII, AndXIV, AndXVI, because the more clustered
satellites are likely to be one coherent group. 
We can only hope that the Hubble Space Telescope will live long enough to
test this, or that future observatories will be able to perform such
measurements.
An alternative will be by combining future adaptive optic imaging with 
earlier Hubble Space Telescope imaging to get proper motions. Water maser 
observations with VLBI have also been used to derive proper motions in the 
LG \citep{brunthaler2005,darling2011}, but this method is likely inapplicable 
to most of M31's low mass satellites because of their lack of gas or star formation.

\subsection{On satellite population models}
The properties of the planes of satellites we find are strongly affected by
the stellar mass model used. 
This simply reflects the fact that we do not understand star formation in
the lowest mass satellites of M31 and MW. 
Any study into the significance of planar structures of satellites, such as
\cite{bahl2013,pawlow2014,ibata2014} will face similar problems. Therefore
future studies investigating disks of faint satellites (i.e. fainter than
classical satellites, i.e. $M_V > -10$) should account for the modelling
uncertainties by exploring various models (even crudely) as we do here. 
In the present paper, our modelling was carried out with simplicity in mind, 
and given the small size of our sample (2 galaxies only, although resulting 
from carefully built constrained initial conditions) it is perilous to assess 
the validity of our models based on so little evidence. Nevertheless, we note 
that the most realistic and significant planes are obtained using the 
${\rm z_{r}}{\rm M_{max}}$ model. Should this be confirmed with additional LG 
simulations, this would  suggest that an accurate modelling of the reionization 
of the LG as we have done here is an important ingredient to reproduce the 
properties of the LG's satellite systems, as was pointed out in \cite{ocvirk2011}.
On the other hand, this problem should not affect studies relying on the
brightest satellites of MW-type galaxies. In particular, the over-abundance
of diametrically opposed co-rotating satellites reported by \cite{ibata2014b}
involves large LMC-type objects, which will come out as the most luminous
satellites in the majority of our models.

\section{Conclusions}
We have searched for planar structure in a high resolution N-body SPH
$\Lambda$CDM simulation of the Local Group, containing a Milky Way - M31
galaxy pair.
We model the satellite populations using five different sets of very simple
recipes, reproducing roughly the variety of models considered in the
literature.
We describe a method for finding the plane containing the maximum of
satellites and validate it on the observed VPoS of M31 found by I13.
The model satellite population spatial distribution is strongly
  dependent on the model prescriptions, and so are the properties of the
  planar configurations we find.

Since the satellite systems of our simulated galaxies may be instrinsically
different from the real M31 (for instance with respect to their radial
distribution), we focus on quantifying the significance of the simulated
planes, as their probability to occur in a random population. This allows us
to quantify the degree of structure, or non-randomness of the satellite
systems, in a manner similar to I13.
Applying this method to our satellite population models we attempt to
compare the simulation to the observations, especially the plane of
satellites of Andromeda (Ibata et al. (2013)) found by PAndAS. We also
consider two alternative volumes (one extended PAndAS and one spherical) in order
to further explore the simulation. We find several cases (a total of seven of
planes which are very unlikely ($<1$\% chances) to be random alignments,
showing that the simulated satellite populations are indeed highly
structured.
Our best maximum plane has 14 of 27 satellites in a plane of 14.1 kpc
dispersion, among which 11 are co-rotating. However, the observed VPoS of
M31 is slightly richer in satellites, has a stronger co-rotation and is
still slightly thinner, which makes it stand out as overall more exceptional
than our simulated planes by a factor ten or more in significance.
The most significant simulated planes tend to be obtained with the ${\rm z_{r}}{\rm M_{max}}$ 
model, highlighting the possibly important role of a realistic description of the inside-out reionization 
of the LG galaxies in investigations of the properties of its low-mass satellite systems.
Most of the simulated planes consist of one coherent group containing about
half of the plane's satellites and forming its backbone, aligning by chance
with another group or several isolated, kinematically unrelated satellites.
This is the main result of this study.
Therefore, although the planes we find are generally dominated by one real structure, 
they are also partly fortuitous and are thus not kinematically coherent structures as a whole:
1/3 to 1/2 of their satellites will fly out of the plane on a short timescale ($\sim 150$ Myr), 
although the main defining group may conserve its alignment and realign by chance with 
another set of satellites.

Provided that the simulated and observed planes of satellites are indeed of
the same nature, our results suggest that the VPoS of M31 is not
a coherent disc and that 1/3 to 1/2 of its satellites must have large proper
motions perpendicular to the plane. We hope that future observational
campaigns will be able to settle this debate.

\section*{Acknowledgements}

This study was performed in the context of the EMMA $(ANR-12-JS05-0001)$ and
LIDAU project $(ANR-09-BLAN-0030)$, funded by the Agence Nationale de la
Recherche (ANR).
The RT computations were performed using HPC resources from
GENCI-[CINES/IDRIS] (Grant 2011-[x2011046667]), on the hybrid queue of
titane at Centre de Calcul Recherche et Technologie, as well as Curie,
during a grand challenge time allocation (project PICON: Photo-Ionisation of
CONstrained realizations of the local group). 
The CLUES simulations were performed at the Leibniz Rechenzentrum Munich
(LRZ) and at the Barcelona Supercomputing Center (BSC). 
AK is supported by the {\it Ministerio de Econom\'ia y Competitividad}
  (MINECO) in Spain through grant AYA2012-31101 as well as the
  Consolider-Ingenio 2010 Programme of the {\it Spanish Ministerio de
    Ciencia e Innovaci\'on} (MICINN) under grant MultiDark CSD2009-00064. He
  also acknowledges support from the {\it Australian Research Council} (ARC)
  grants DP130100117 and DP140100198. He further thanks Stella for l'idole
  des jaunes.
The author thanks C. Scannapieco for precious hints dispensed in the initial
phase of the project, as well as the CLUES collaborators for useful
discussions. 
The author thanks D. Munro for freely distributing his Yorick programming
language\footnote{\label{yo} http://www.maumae.net/yorick/doc/index.html},
and its yorick-gl extension which was used in the course of this study.

\bibliography{biblio}

\newpage

\end{document}